\documentclass[12pt]{article}
\usepackage{latexsym}
\usepackage{amssymb}
\usepackage{amsfonts}
\usepackage{pslatex}
\usepackage{graphicx}

\catcode`\@=11
\def\numberbysection{\@addtoreset{equation}{section}
        \def\theequation{\thesection.\arabic{equation}}}

\def\be{\begin{equation}}
\def\ee{\end{equation}}
\def\ba{\begin{eqnarray}}
\def\ea{\end{eqnarray}}

\def\ov{\overline}

\def\Z{\mathbb{Z}}

\def\nl{\nonumber \\}

\def\ra{\rangle}

\def\la{\langle}
\def\de{\partial}

\def\Tr{{\rm Tr}}
\def\dag{\dagger}

\def\G{\Gamma}
\def\D{\Delta}
\def\d{\delta}

\def\eps{\varepsilon}

\def\th{\theta}
\def\k{\kappa}
\def\l{\lambda}
\def\L{\Lambda}

\def\p{\pi}
\def\P{\Pi}

\def\w{\omega}

\def\B{{\bf B}}

\numberbysection

\begin{document}
\begin{titlepage}
\begin{center}

\hfill  \quad DFF 433/06/06 \\
\hfill  \quad NSF-KITP-06-78 \\

\vspace{1cm} 

{\Large \bf Jain States } \\ 
\vspace{.4cm}
{\Large \bf in a Matrix Theory  }\\ 
\vspace{.4cm}
{\Large \bf of the Quantum Hall Effect }\\ 

\vspace{1cm}

Andrea CAPPELLI ${}^{a,b}$,\ \ Ivan D. RODRIGUEZ ${}^a$ \\
\bigskip
{\em ${}^a$ I.N.F.N. and Dipartimento di Fisica,}\\
{\em  Via G. Sansone 1, 50019 Sesto Fiorentino - Firenze, Italy} \\
\medskip
{\em ${}^b$ Kavli Institute of Theoretical Physics, 
University of California, }\\
{\em Santa Barbara, CA 93106 USA}
\end{center}
\vspace{.5cm}
\begin{abstract}
The U(N) Maxwell-Chern-Simons matrix gauge theory is proposed as an
extension of Susskind's noncommutative approach.
The theory describes D0-branes, 
nonrelativistic particles with matrix coordinates and gauge symmetry,
that realize a matrix generalization of the quantum Hall effect.
Matrix ground states obtained
by suitable projections of higher Landau levels are found to
be in one-to-one correspondence with
the expected Laughlin and Jain hierarchical states.
The Jain composite-fermion construction follows by gauge invariance
via the Gauss law constraint.
In the limit of commuting, ``normal'' matrices
the theory reduces to eigenvalue coordinates that describe
realistic electrons with Calogero interaction.
The Maxwell-Chern-Simons matrix theory improves
earlier noncommutative approaches and could 
provide another effective theory of the fractional Hall effect.
\end{abstract}

\vfill
\end{titlepage}
\pagenumbering{arabic}


\section{Introduction}
In this paper we continue the study initiated in \cite{cr}
of noncommutative and matrix gauge theories \cite{NCFT} \cite{mtheory}
that could describe the fractional quantum Hall effect \cite{prange}
as effective (non-relativistic) field theories\footnote{
We refer to \cite{cr} for a more extensive introduction
to noncommutative theories of the quantum Hall effect 
and the account of earlier literature.}.
In this approach, we aim at finding a consistent theoretical
framework encompassing and completing the well-established 
phenomenological theories of Laughlin \cite{laugh} and Jain \cite{jain},
as well as the low-energy effective theories of edge excitations
\cite{edgecft} \cite{winf}\cite{sakita}\cite{jainedge}. 
Our setting is formally rather different from that of
non-relativistic fermions coupled to 
the Chern-Simons interaction developed by Fradkin and Lopez 
\cite{fradkin} and others \cite{cftheories}, 
but there are some analogies in the physical interpretation.

The use of noncommutative and matrix theories was 
initiated by Susskind \cite{susskind}
who observed that two-dimensional semiclassical incompressible fluids 
in strong magnetic fields can be described by the non-commutative
Chern-Simons theory in the limit of 
small noncommutative parameter $\theta$, corresponding to high density.
Afterwards, Polychronakos extended the theory to describe a finite
droplet of fluid, and obtained the U(N) matrix gauge theory
called Chern-Simons matrix model \cite{poly1}.
In this theory, the noncommutativity is realized in terms
of two Hermitean matrix coordinates,  
$[X_1(t),X_2(t)]\ =\ i\ \th\ {\rm I}_N$, and $\th$ corresponds to
a uniform background field\footnote{
Here, ${\rm I}_N$ denotes the traceless $NxN$ pseudo-identity matrix 
defined in Section 2.}.
Polychronakos analyzed the classical droplet solutions and its excitations and 
showed a very interesting relation with the one-dimensional Calogero model,
whose particle have coordinates $x_a$, $a=1,\dots,N,$ corresponding
to the eigenvalues of one matrix, say $X_1$.

Several authors \cite{poly3} \cite{park} \cite{heller} \cite{karabali}
\cite{hansson} \cite{hansson2} \cite{leigh} \cite{cr}
further analyzed the 
Chern-Simons matrix model trying to connect it to the physics
of the fractional Hall effect, most notably the Laughlin states and its
quasi-particle excitations. 
The allowed filling fractions of quantum fluids were found to be, 
$\nu=1/(\B\th +1)=1/(k+1)$, where $\B$ is the external magnetic 
field and $\B\th$ is integer quantized by gauge invariance \cite{nair}.

As in any gauge theory, the Gauss-law condition requires
the states to be singlets of the U(N) gauge group.
In matrix theories the singlet ground state wave function 
in eigenvalue coordinates  $x_a$ contains the Vandermonde factor,
$\D(x)=\prod_{a<b}(x_a-x_b)$, that indicates
the relation with one-dimensional fermions \cite{mehta}.
In the Chern-Simon matrix model, 
the contribution of the $\th$ background forces the states 
to carry a specific representation of the U(N) gauge group:
the non-trivial ground state wave function is found to be a power of the
Vandermonde, $\prod_{a<b}(\l_a-\l_b)^{k+1}$, in terms of the
complex eigenvalues $\l_a$, $a=1,\dots,N$, of $X=X_1+i\ X_2$ 
\cite{heller}.
This is actually the Laughlin wave function at filling
$\nu=1/(k+1)$ \cite{laugh}, upon interpreting the eigenvalues as the
coordinates of N planar electrons in the lowest Landau level.
Therefore, the celebrated Laughlin state was shown to be the
exact ground state of the matrix theory that is completely determined by
gauge invariance. 
This is the nicest result obtained in 
the matrix (noncommutative) approach. 

In spite of these findings, the Chern-Simons matrix model so far
presented some difficulties that limited its applicability
as a theory of the fractional Hall effect \cite{hansson2}:
\begin{itemize}
\item
The Chern-Simons matrix model does not possess quasi-particle excitations,
only quasi-holes can be realized \cite{poly1}.
\item
The Jain states with the filling fractions, 
$\nu=m/(mk+1)$, $m=2,3,\dots$,  cannot be realized in the theory, 
even including more boundary terms \cite{poly3}.
\item
Even if the Laughlin wave function is obtained, the measure of
integration differs from that of electrons in the lowest Landau level,
owing to the noncommutativity of matrices \cite{karabali}.
As shown in Ref.\cite{cr}, 
the ground state properties of the matrix theory and of
the Laughlin state agree at long distances but differ microscopically.
\item
Owing to the inherent noncommutativity, 
it is also difficult to match matrix observables
with  electron quantities of the quantum Hall effect \cite{hansson}.
\end{itemize}

In this paper, we show that some of these problems can be overcome by 
upgrading the Chern-Simons model to the Maxwell-Chern-Simons matrix theory.
This includes an additional kinetic term quadratic in time derivatives and
the potential $V = - g \Tr \left[ X_1,X_2 \right]^2$,
parametrized by the positive coupling constant $g$.
All the terms in the action are fixed by the gauge principle
because they are obtained by dimensional reduction of
the three-dimensional Maxwell-Chern-Simons theory.
The matrix theory has been discussed in the literature of string
theory as the low-energy effective theory of a stack of N D0-branes on
certain higher-brane configurations \cite{dzero}; in particular, D0-branes
have been proposed as fundamental degrees of freedom in string 
theory \cite{mtheory}.

In section two, we introduce the  Maxwell-Chern-Simons matrix theory, quantize
the Hamiltonian and discuss the Gauss-law constraint on physical states.
The Hamiltonian contains a kinetic term, involving matrix coordinates,
$X_1,X_2$, and conjugate momenta, $\P_1,\P_2$,
and the potential $V$ parametrized by $g$.
The kinetic term realizes a matrix analogue of the Landau levels:
it decomposes into $N^2$ copies of the Landau Hamiltonian for the
``particles'' whose coordinates are the matrix entries,
$X^1_{ab}, X^2_{ab}$, $a,b=1,\dots,N$.
The energy scale is set by the Landau-level gap $\B/m$.
The theory has an interesting phase diagram in terms
of the parameters, $\B/m$ for the kinetic energy and $g$ for the potential,
while  $\B\th=k \in \Z$ is
fixed by the choice of density, i.e. of filling fraction.

In section three we discuss the theory for $g=0$ and find the physical 
gauge-invariant states by solving the Gauss law.
In the zero-energy sector, corresponding to the lowest Landau level, 
the theory reduces to the earlier Chern-Simons matrix model and 
exhibits the Laughlin ground states.
General $E>0$ gauge-invariant states are SU(N) singlets that resemble Slater
determinants of N fermions: therefore, we can set up a 
pseudo-fermionic Fock space description of matrix states, but find
higher degeneracies due to matrix noncommutativity.
We introduce a series of projections in the theory that gradually
reduce the number of available states:
in the $m$-projected theory, $m=2,3,\dots$, we only allow states
belonging to the lowest $m$ matrix Landau levels.
Each of the projected theories
possesses non-degenerate ground states corresponding to the
uniform filling of allowed levels, that
remarkably match the pattern envisaged by Jain in his
phenomenological theory of ``composite fermions'' \cite{jain}.
These states have the 
expected filling fraction $\nu=m/(mk+1)$, with $k=\B\th$, and
admit quasi-particle excitations. 

To summarize, the Maxwell-Chern-Simons matrix theory at $g=0$
can be suitably truncated to display non-degenerate ground states that 
are matrix generalizations of Jain's composite-fermion wave functions  
in the fractional Hall effect.
This is the main result of this paper.

In section four, we switch on the  potential,
$V = - g\ \Tr \left[ X_1,X_2 \right]^2$.
For $g\to\infty$, the matrices are
forced to commute among themselves, and thus can be simultaneously
diagonalized by unitary transformation\footnote{
The complex matrix $X=X_1+iX_2$ commutes with its adjoint
and is called ``normal'' \cite{wiegmann}.}: 
$X^i=U\ \L^i U^\dagger$, $\L^i={\rm diag}(\l^i_a)$, $i=1,2$, $a=1,\dots,N$.
Therefore, $N(N-1)/2$ physical, non-diagonal matrix components
are projected out.
The momenta $\P_1,\P_2$ remain noncommutative, but
their non-diagonal parts are completely determined by
the Gauss-law condition \cite{poly97}\cite{park}.
Once put back into the Hamiltonian, these terms
induce a two dimensional Calogero interaction among the remaining
degrees of freedom $\l^i_a$.
The complete reduction to eigenvalues allows them to be interpreted
as coordinates of N electrons: their Calogero interaction
$O(k^2/r^2)$ is a legitimate replacement of the Coulomb potential
$O(e^2/r)$, because 
the universal properties of incompressible Hall fluids are
rather independent of the specific form of potential, for large values of $\B$
\cite{laugh}\cite{haldane}\cite{jain}.
Therefore, for $g\to\infty$ the Maxwell-Chern-Simons matrix theory
definitely describes the fractional quantum Hall effect.

We remark that the $g=\infty$ theory cannot easily be solved:
as in the original problem, there are no small parameters to expand
and the non-perturbative gap is due to particle interaction.
Nevertheless, we point out that the $g=\infty$  theory can be
understood by relating it to the solvable $g=0$ theory, following the
deformation of $g=0$ states caused by the potential
$V = - g\ \Tr \left[ X_1,X_2 \right]^2$.

In section five, we discuss the phase diagram
of the Maxwell-Chern-Simons theory as a function of $g$ and $\B/m$.
We note that the non-degenerate matrix Jain states found at $g=0$ 
are not eigenstates of the potential $V$ and  evolve
into other unknown states for $g>0$.
Nevertheless,  the $g=0$ matrix states evaluated
for diagonal $X,\ov{X}$ expressions (i.e. at $g=\infty$) become 
Slater determinants that are exactly equal to 
the (unprojected) Jain wave functions \cite{jain}: 
extensive numerical results
\cite{laugh}\cite{haldane}\cite{jain}\cite{numeric}
indicate that these states are very close to the exact energy eigenstates 
for several short-range interactions including $1/r^2$.

This fact let us to conjecture that the matrix ground states
found at $g=0$, corresponding to the Laughlin and Jain series, 
remain non-degenerate for all $g$ values: 
namely, that there are no phase changes in the theory for the density values
characteristic of  $g\sim 0$ non-degenerate states.
 
Although in this paper we cannot provide a proof
of this conjecture, we find it rather compelling
and worth describing in some detail.
We remark that approximate methods can treat perturbatively or 
self-consistently the $\Tr \left[ X_1,X_2 \right]^2$
potential around the $g=0$ ground states, owing to their non-degeneracy.
Our approach here is similar in spirit to that of the Lopez-Fradkin 
theory \cite{fradkin}, where degeneracy was removed 
by adding the Chern-Simons interaction.
If the $V$ potential can be efficiently treated,  
the Maxwell-Chern-Simons matrix theory could provide
a new effective theory of the fractional Hall effect,
where the Laughlin and Jain ground states naturally 
appear at $g=0$, owing to matrix gauge invariance, and get modified
as $g\to\infty$ while remaining in the same universality class.


\section{U(N) Maxwell-Chern-Simons matrix gauge theory}

We start by discussing the canonical analysis of the Maxwell-Chern-Simons 
matrix theory \cite{poly97}\cite{park}, 
in presence of the uniform background $\th$ and the
``boundary'' term of Ref.\cite{poly1}.
The theory involves three time-dependent $N\times N$ Hermitean matrices,
$X_i(t)$, $i=1,2$ and $A_0(t)$, and the auxiliary complex 
vector $\psi(t)$: it is defined by the action,
\ba
S&=&\int \ dt\ \Tr\left[
\frac{m}{2}\left(D_t\ X_i \right)^2 \ +\ 
\frac{\B}{2} \eps_{ij} \ X_i\ D_t \ X_j \ +\ 
\frac{g}{2} \left[ X_1,X_2\right]^2 \ +\ \B\th\ A_0
\right]
\nl
& &-i\ \int \psi^\dag\ D_t \psi \ .
\label{mcs-action}
\ea
The form of the covariant derivatives is:
$D_t X_i=\dot{X}_i -i \left[A_0,X_i\right]$ and
$D_t \psi = \dot{\psi}-i A_0\psi$.

For $m=0$, the theory reduces to the Chern-Simons matrix model \cite{susskind},
fully analyzed in Ref. \cite{poly1}: indeed, in this limit the potential 
$\Tr \left[X_1,X_2\right]^2$ becomes a constant for all states 
(see section 2.2).
Hereafter we set $m=1$ and measure dimensionful constants accordingly.
Under U(N) gauge transformations: $X_i\ \to\ U X_i U^\dag$,
$A_0\ \to\ U \left(A_0-id/dt \right) U^\dag$, 
and $\psi\ \to\ U\psi$, the action changes by
a total derivative, such that invariance under large gauge
transformations requires the quantization, $\B\th=k\in \Z$,
as in the case of the Chern-Simons model \cite{nair}.

The canonical momenta are the following Hermitean matrices:
\be
\P_i\equiv \frac{\d S}{\d \dot{X}^T_i}\ = \  
D_t X_i -\frac{\B}{2} \eps_{ij} X_j \ ,
\label{pi-def}
\ee
and  $\chi=\d S/\d \dot{\psi}= -i\psi^\dag$,
After Legendre transformation on these variables, one finds the Hamiltonian:
\be
H \ =\ \Tr \left[
\frac{1}{2}\left(\P_i +\frac{\B}{2}\ \eps_{ij}\ X_j \right)^2 \ -\ 
\frac{g}{2} \left[X_1,X_2 \right]^2\ 
\right]\ . 
\label{mcs-ham}
\ee

The variation of $S$ w.r.t. the non-dynamical field $A_0$ gives the 
Gauss-law constraint; its expression in term of coordinate and momenta 
reads:
\be
G\ = \ 0\ , \qquad\quad
G\ =\ i\ \left[X_1,\P_1\right]\ +\ i\ \left[X_2,\P_2\right]\ 
-\ \B\th\ {\rm I}\ +\ \psi\otimes \psi^\dag \ ,
\label{mcs-gauss}
\ee
where I is the identity matrix.
By taking the trace of $G$, one fixes the norm of the auxiliary vector
$\psi$,
\be
\Tr\ G\ =\ 0\quad \longrightarrow\ \Vert\psi\Vert^2=\B\th N=kN\ .
\label{psi-norm}
\ee
We note that the auxiliary vector has vanishing Hamiltonian 
and trivial dynamics, $\psi(t)=\psi(0)={\rm const.}$:
as in the Chern-Simons model, it is necessary to represent the Gauss law on
finite-dimensional matrices that have traceless commutators \cite{poly1}.
In a gauge in which all $\psi$ components vanish but one,
the term $\B\th \ {\rm I}\ -\psi\otimes \psi^\dag$ in (\ref{mcs-gauss})
is replaced by the traceless ``identity'',
$ \B\th \ {\rm I}_N$, $\ {\rm I}_N=diag(1,\cdots,1,1-N)$.
At the quantum level, the operator $G$ generates 
U(N) gauge transformations of $X_i$
and $\psi$, and requires the physical states to be U(N) singlets subjected
to the additional condition (\ref{psi-norm}) counting the number of 
$\psi_a$ components.


\subsection{Covariant quantization}

We now quantize all the $2N^2$ matrix degrees of freedom $X^i_{ab}$ and
later impose the Gauss law as a differential condition on wave functions
in the Schroedinger picture.
The Hamiltonian (\ref{mcs-ham}) for $g=0$ is quadratic and easily solvable:
the sum over matrix indices decomposes into $N^2$  identical terms
that are copies of the Hamiltonian of Landau levels \cite{winf}.
To see this, introduce the matrix:
\be
A=\frac{1}{2\ell}\left( X_1 +i\ X_2 \right)\ +\
\frac{i\ell}{2}\left( \P_1 +i\ \P_2 \right)\ ,
\qquad 
\label{a-def}
\ee
and its adjoint $A^\dag$, involving the constant 
$\ell=\sqrt{2/\B}$ called ``magnetic length''. 

The quantum commutation relations following from (\ref{pi-def}) are,
\be
\left[\left[ X^i_{ab}, \P^j_{cd} \right]\right]\ = 
\ i\ \d^{ij}\ \d_{ad}\ \d_{bc}\ ,
\qquad\qquad
\left[\left[ \psi^\dag_a, \psi_b \right]\right]\ = \ \d_{ab}\ ;
\label{can-comm}
\ee
they are written using double brackets to distinguish them 
from classical matrix commutators.
The canonical commutators imply the following relations of $N^2$ 
harmonic oscillators:
\be
 \left[\left[A_{ab},A^\dag_{cd} \right]\right] \ = \ \d_{ad}\ \d_{bc}\ \ ,
\qquad\qquad
\left[\left[A_{ab},A_{cd}\right]\right] \ =\  0\ .
\label{a-comm}
\ee
Note that $A^\dag$ is the adjoint of $A$ both as a matrix and a
quantum operator.
The Hamiltonian can be expressed in term of $A$ and $A^\dag$ as
follows:
\be
H \ =\ \B\ \Tr \left( A^\dag \ A \right)\ +\ \frac{\B}{2} N^2\ -
\frac{g}{2} \Tr \left[X_1,X_2 \right]^2\  . 
\label{a-ham}
\ee
In the term $\Tr(A^\dag A)=\sum_{ab} A^\dag_{ab} A_{ba} $ one recognizes
$N^2$ copies of the Landau level Hamiltonian corresponding
to $N^2$ two-dimensional ``particles'' with phase-space coordinates,
$\{\P^i_{ab},X^i_{ab}\}$, $a,b=1,\dots,N$, $i=1,2$.

The one-particle state are also characterized by another set of
independent oscillators corresponding to angular momentum excitations 
that are degenerate in energy and thus occur within each Landau
level. To find them, introduce the matrix,
\be
B=\frac{1}{2\ell}\left( X_1 -i\ X_2 \right)\ +\
\frac{i\ell}{2}\left( \P_1 -i\ \P_2 \right)\ ,
\qquad 
\label{b-def}
\ee
and its adjoint $B^\dag$. They obey:
\be
 \left[\left[B_{ab},B^\dag_{cd} \right]\right] \ = \ \d_{ad}\ \d_{bc}\ \ ,
\qquad\qquad
\left[\left[B_{ab},B_{cd}\right]\right] \ =\  0\ ,
\label{b-comm}
\ee
and commute with all the $A_{ab},A^\dag_{ab}$. 

The total angular momentum of the $N^2$ ``particles'' can be
written in the U(N) invariant form
\be
J\ =\ \Tr\left( X_1\ \P_2\ - \ X_2\ \P_1 \right)\ = \
 \Tr\left( B^\dag B\ - \ A^\dag A\right)\ .
\label{j-def}
\ee
Therefore, the $B$ oscillators count the angular momentum excitations
of the particles within each Landau level. In conclusion, the
$g=0$ theory exactly describes $N^2$ free particles in the
Landau levels. In section three we shall discuss 
the effect of gauge symmetry that selects the subset of multi-particle states
obeying the Gauss law  $G=0$ (\ref{mcs-gauss}).


\subsection{Projection to the lowest Landau level and Chern-Simons 
matrix model}

For large values of the magnetic field $\B$, one often considers
the reduction of the theory to the states in lowest Landau level that have
vanishing energy (\ref{a-ham}), i.e. obey $A_{ab}=0$ $\ \forall a,b$. 
All higher levels can be projected out by imposing the
constraints $A=A^\dag=0$, that can be written classically (cf (\ref{a-def})):
\be
\P_2=\frac{\B}{2}\ X_1\ , \qquad\qquad \P_1=-\frac{\B}{2}\ X_2 \ ,
\label{lll-proj}
\ee
corresponding again to vanishing kinetic term in the Hamiltonian 
(\ref{mcs-ham}).

In this projection, two of the four phase space coordinates per particle
are put to zero: if we choose them to be $\P_1,\P_2$, the remaining 
variables $X_1,X_2$ become canonically conjugate.
This can also be seen from the action (\ref{mcs-action}), 
because the kinetic term  $m\left(D_i X\right)^2$
vanishes and one is left with the Chern-Simons term implying the
identification of one coordinate with a momentum \cite{dunne} \cite{winf}.

Upon eliminating $\P_1,\P_2$, the Gauss law (\ref{mcs-gauss}) becomes:
\be
G \ = \  - i\B\ \left[X_1,X_2\right] - \B\th +\psi\otimes\psi^\dag\ ;
\label{cs-gauss}
\ee
namely, it reduces to the noncommutativity condition of the
Chern-Simon matrix model, with action \cite{poly1},
\be
S_{CSMM}=\int \ dt\ \Tr\left[
\frac{\B}{2} \eps_{ij} \ X_i\ D_t \ X_j \ +\ \B\th\ A_0
\right]
-i\ \int \psi^\dag\ D_t \psi \ .
\label{cs-action}
\ee
Moreover, the potential term in the Hamiltonian (\ref{mcs-ham})
becomes a constant on all physical
states verifying $G=0$: one finds, using the normalization (\ref{psi-norm}),
\be
-\ \Tr\left[X_1,X_2\right]^2\ =
\ \Tr \left(\th\ {\rm I} -\frac{1}{\B}\psi\otimes \psi^\dag\right)^2\ 
= \ \th^2\ N(N-1) \ .
\label{cs-pot}
\ee
In conclusion, the Hamiltonian (\ref{mcs-ham}) reduces to a constant,
i.e. it vanishes.
This completes the proof that the Maxwell-Chern-Simons matrix theory
projected to the lowest Landau level  is equivalent to the
previously studied Chern-Simons matrix model.
As shown in Ref.\cite{poly1}, this theory is the finite-dimensional
regularization of the noncommutative Chern-Simons theory \cite{susskind}.
In particular, the Gauss law (\ref{cs-gauss}) can be rewritten,
\be
\left[ X_1,X_2\right]\ =\ i\ \th \ {\rm I}_N\ , 
\qquad {\rm I}_N =\ diag \left(1,\cdots,1,1-N\right)\ ,
\label{nc-rel}
\ee
in the gauge in which $\psi$ has only one non-vanishing 
component (the N-th one).
Equation (\ref{nc-rel}) expresses the coordinate noncommutativity 
in the limit $N\to\infty$ \cite{ncgeom}.


\section{Physical states at $g=0$ and the Jain composite-fermion 
correspondence}

In this section we are going to solve the Gauss law condition 
(\ref{mcs-gauss}) and find the gauge invariant states.
We first recall the form of physical states in the lowest 
Landau level, equal to those of the Chern-Simons matrix model already
found in Ref.\cite{poly1}\cite{heller} (section 3.1), 
and then discuss the general physical states (sections 3.3,3.4).
We introduce the complex matrices,
\ba
X &=& X_1+i\ X_2\ , \qquad\qquad \ov{X}=X_1 -i\ X_2\ ,
\nl
\P &=& \frac{1}{2}\left(\P_1 -i\ \P_2\right)\ ,
\qquad \ov{\P}=\frac{1}{2}\left(\P_1 +i\ \P_2\right)\ ,
\label{b-mat}
\ea
and use the bar for denoting the Hermitean conjugate of
classical matrices, keeping the dagger for the quantum adjoint.
We set the magnetic length to one, i.e. $\B=2$.
The wave functions of the Maxwell-Chern-Simons theory 
take the form,
\be
\Psi\ = \ e^{-\Tr\left(\ov{X} X \right)/2 -\ov{\psi} \psi/2}\ 
\Phi(X,\ov{X},\psi) \ .
\label{wf-def}
\ee
For energy and angular momentum eigenstates, 
the function $\Phi$ in (\ref{wf-def}) 
is a polynomial in the matrices $X,\ov{X}$ and the auxiliary field $\psi$.
The integration measure reads:
\be
\left\la \Psi_1 \vert \Psi_2 \right\ra \ = \
\int {\cal D}X {\cal D}\ov{X}\ {\cal D}\psi {\cal D}\ov{\psi}\  
e^{-\Tr \ov{X} X -\ov{\psi} \psi}\ 
\Phi^*_1(X,\ov{X},\psi)\ \Phi_2(X,\ov{X},\psi)\ .
\label{int-meas}
\ee

The operators $A_{ab},B_{ab}$ and $\psi_a,\psi^\dag_b$
characterizing the Hilbert space (cf. section 2.2)
become differential operators acting on wave functions:
\ba
A_{ab} &=& \frac{1}{2}\ X_{ab} + i \ \ov{\P}_{ab} \ = \ 
\ \frac{X_{ab}}{2} + \ \frac{\de\ \ }{\de \ov{X}_{ba}}\ ,
\qquad\quad
A_{ab}^\dagger \ = \ 
\ \frac{\ov{X}_{ab}}{2} - \ \frac{\de\ \ }{\de X_{ba}}\ ,
\nl
B_{ab} &=& \frac{\ov{X}_{ab}}{2} + \ \frac{\de\ \ }{\de X_{ba}}\ ,
\qquad\quad
B_{ab}^\dagger = \frac{X_{ab}}{2} - \ \frac{\de\ \ }{\de \ov{X}_{ba}}\ ,
\qquad\quad
\psi^\dag_a\ =\ \frac{\de\ \ }{\de \psi_{a}}\ .
\label{diff-op}
\ea
Correspondingly, the Gauss law condition (\ref{mcs-gauss}) becomes:
\ba
&& G_{ab}\ \Psi_{\rm phys}(X,\psi) \ =\  0\ , \nl
&& G_{ab} = \sum_c \left( 
X_{ac} \ \frac{\de\ \ }{\de X_{bc}} - X_{cb}\ \frac{\de\ \ }{\de X_{ca}} 
+ \ov{X}_{ac}\ \frac{\de\ \ }{\de \ov{X}_{bc}} - 
\ov{X}_{cb}\ \frac{\de\ \ }{\de \ov{X}_{ca}} 
\right) 
- k\ \d_{ab} + \psi_a\ \frac{\de\ \ }{\de \psi_b} \ . 
\label{diff-gauss}
\ea 
This operator acting on wave functions performs an infinitesimal gauge
transformation of its variables: $X,\ov{X},\psi$. Note that
the expression of $G_{ab}$ in (\ref{diff-gauss}) was normal ordered for this
to obey the U(N) algebra \cite{poly1}.

The action of the angular momentum (\ref{j-def}) 
on the polynomial part of wave functions is,
\be
\sum_{ab} \left( 
X_{ab} \ \frac{\de\ \ }{\de X_{ab}} - 
\ov{X}_{ab}\ \frac{\de\ \ }{\de \ov{X}_{ab}}
\right)\ 
\Phi(X,\ov{X},\psi) \ = \ {\cal J} \ \Phi(X,\ov{X},\psi) \ .
\label{j-diff}
\ee
The eigenvalue ${\cal J}$ is just the total number of 
$X$ matrices occurring in $\Phi$ minus that of $\ov{X}$. 
For states with constant density\footnote{
See Refs.\cite{poly1}\cite{cr} for the definition of the density in 
the matrix theory.},
the angular momentum measures the extension of the ``droplet of fluid'',
such that we can associate a corresponding
filling fraction $\nu$ by the formula,
\be
\nu\ =\ \lim_{N\to\infty}\ \frac{N(N-1)}{2{\cal J}}\ .
\label{nu-def}
\ee

In a physical system of finite size, one can control
the density of the droplet, i.e. the angular momentum, by adding
a confining potential $V_C$ to the Hamiltonian:
\be
H \ \to \ H\ +\ V_C \ =\ H\ +\ \w\ \Tr\left(B^\dag\ B  \right) \ .
\label{conf-pot}
\ee
and tune its strength $\w$. This potential is diagonal on all states
and becomes quadratic in the lowest Landau level, $V_C\to \w \Tr (\ov{X}X)$. 
Typical values for $\w$
will be of order $\B/N$, that do not destroy the Landau-level structure
but give a small slope to each level.


\subsection{The Laughlin wave function}

The multi-particle states in the lowest Landau level obey the conditions
$A_{ab}\Psi=0$, $\forall\ a,b$: 
using (\ref{diff-op}) and (\ref{wf-def}), they read, 
\be
A_{ab}\ \Psi\ =\ 0\ \quad\longrightarrow\quad
\frac{\de\ \ }{\de \ov{X}_{ab}}\ \Phi(X,\ov{X},\psi) \ =\ 0
\ ,\qquad \forall\ a,b \ .
\label{lll-cond}
\ee
They imply that the polynomial $\Phi$ does not contain any $\ov{X}_{ab}$,
i.e. is a analytic function of the $X_{ab}$ variables, 
as in the ordinary Landau levels.
The gauge invariant states should
be given by polynomials $\Phi(X,\psi)$ that satisfy the Gauss law
(\ref{mcs-gauss},\ref{diff-gauss}): namely, they are U(N) singlets 
that contain $Nk$ copies of the  $\psi$ vector
(from $\Tr G=0$ in (\ref{diff-gauss})).
These solutions were already found in Ref.\cite{heller}
for the equivalent Chern-Simons matrix model: let us recall the argument.
One can form a polynomial of an arbitrary number of $X_{ab}$
and saturate the indices with $\psi^a$'s and the  U(N) invariant
N-component epsilon tensor. 
For $k=1$, one obtain the states:
\be
\Phi_{\{n_1,\dots,n_N\}}\left(X,\psi\right)\ = \ 
\eps^{a_1\dots a_N}\ \left(X^{n_1}\psi\right)_{a_1}\cdots
\left(X^{n_N}\psi\right)_{a_N} \ ,\qquad 0\le n_1<n_2<\cdots <n_N\ ,
\label{HVR-states}
\ee
for any set of positive ordered integers $\{n_i\}$.
The ground state in the confining potential $\Tr( X X^\dag)$
is given by the closest packing $\{0,1,\dots,N-1\}$ that has the lowest
angular momentum, i.e. lowest degree in $X$.
Solutions for $k\neq 1$ are obtained by multiplying
$k$ terms (\ref{HVR-states}), leading to the expressions,
$\Phi_{\{n^1_1,\dots,n^1_N\}\cdots\{n^k_1,\dots,n^k_N\}}$.

As shown in Ref.\cite{heller}, an equivalent basis for these
states is given by:
\ba
\Phi \left(X,\psi\right) &=& \sum_{\{m_k\}}\ \Tr \left(X^{m_1}\right)\cdots
\Tr \left(X^{m_k}\right) \ \Phi_{k,\ gs}\ ,\nl
\Phi_{k,\ gs} &=& \left[\eps^{a_1\dots a_N} \ 
\psi_{a_1}\ \left(X\psi\right)_{a_2}\cdots \left(X^{N-1}\psi\right)_{a_N}
\right]^k\ ,
\label{HVR-bose}
\ea
which is factorized into the ground state $\Phi_{k,\ gs}$ and
the ``bosonic'' powers of $X$, $\Tr \left( X^{m_i}\right)$,
with positive integers $\{m_1,\dots,m_k\}$ now unrestricted.

Let us now suppose that we diagonalize the complex matrix $X$ by the 
similarity transformation:
\ba
X &=& V^{-1}\ \L\ V\ ,\qquad\qquad \L\ = \ diag(\l_1,\dots,\l_N)\ , \nl
\psi &=& V^{-1}\ \phi\ ,
\label{Bogo-trans}
\ea
where the $\l_a$'s are complex numbers.
The dependence on $V$ and $\phi$ factorizes in the ground state wave function 
and the powers of eigenvalues 
make up the Vandermonde determinant $\D(\l)=\prod_{a<b}(\l_a-\l_b)$,
as follows: 
\ba
\Phi_{k,\ gs}\left(\L,V,\psi\right) &=& \left[ \eps^{a_1\dots a_N} \ 
\left(V^{-1}\phi\right)_{a_1}\ 
\left(V^{-1}\L\phi\right)_{a_2}\cdots 
\left(V^{-1}\L^{N-1}\phi\right)_{a_N} \right]^k\ \nl
&=&\left[\left(\det V\right)^{-1} \ 
\det \left( \l_a^{i-1}\ \phi_a \right) \right]^k \nl
&=& \left(\det V\right)^{-k} 
\ \prod_{1\le a< b\le N}\left(\l_a-\l_b\right)^k\  
\left(\prod_c \ \phi_c\right)^k\ .
\label{Laugh-wf}
\ea
The central piece is indeed the Laughlin wave function for the ground state of
the Hall effect \cite{laugh}, with eigenvalues as
electron coordinates \cite{heller} \cite{karabali}. 
The value of the filling fraction is:
\be
\nu\ =\ \frac{1}{k+1}\ ,
\label{nu}
\ee
and is renormalized from the classical value (\ref{nu-def}),
$\nu=1/\B\th=1/k$, because the wave functions acquire
one extra factor of $\D(\l)$ from the integration measure (\ref{int-meas})
reduced to the eigenvalues \cite{karabali}. 
Physical values of $\nu$ correspond to even $k$'s and
antisymmetric wave functions.
The factorized dependence on $V$ and $\phi$ in (\ref{Laugh-wf}) 
is the same for all the states (\ref{HVR-bose}), because they do not
occur in the power sums $\Tr\left(X^r \right)= \sum_a\l_a^r$.
The $\psi$ dependence can be integrated out, while the dynamics
of the additional degrees of freedom described by $V$ will be discussed
later (see section four). 

As anticipated in the introduction, Eq.(\ref{Laugh-wf}) 
is the most intriguing result obtained in the noncommutative approach and
the Chern-Simons matrix model: that of deriving
the Laughlin wave function from gauge invariance in a matrix theory.
Furthermore, Susskind's semiclassical analysis \cite{susskind} 
showed that, in the limit $\th\to 0$,
this matrix state  describes an incompressible fluid in high magnetic fields, 
with density,
\be
\rho_o\ = \ \frac{1}{2\pi \th} \ ,
\label{rho-def}
\ee
in agreement with the earlier identification of the filling fraction.

Let us discuss the excitations of the matrix Laughlin states.
Multiplying the 
wave function by polynomials of $\Tr(X^r)$ as in (\ref{HVR-bose}), 
we find  states with $\D {\cal J} =r$. These
are the basis of holomorphic excitations over the Laughlin state.
For $r=O(1)$, their energy given by the boundary potential, 
$\D E=\w \D {\cal J} = O(r\ \B/N)$ is very small:
they are the degenerate edge excitations of the droplet of 
fluid described by conformal field
theories \cite{edgecft}\cite{winf}\cite{sakita}\cite{jainedge}.

More interesting hereafter are the analogues of the quasi-hole and
 quasi-particle excitations of the Laughlin state, that
are gapful localized  density deformations.
The quasi-hole is realized
by moving one electron from the interior of the Fermi surface
to the edge, causing $\D {\cal J} =O(N)$ and thus a finite gap
$\D E=O(\B)$. 
Its realization in the matrix theory is for example given by the state 
$\Phi_{\{n_1,\dots,n_N\}}$ in Eq.(\ref{HVR-states}), with 
$\{n_1,n_2,\cdots,n_M\}=\{1,2,\cdots,N\}$.
On the other hand, the quasi-particle excitation cannot be 
realized in the Chern-Simons matrix model \cite{poly1}.


\subsection{The Jain composite-fermion transformation}

The result (\ref{rho-def}) can actually be interpreted in the language of
the Jain composite-fermion transformation \cite{jain}.
According to Jain, a system of electrons with inverse filling fraction
parametrized by:
\be
\frac{1}{\nu} = \frac{\B}{2\pi\rho_o} = \frac{1}{m}+k \ ,\qquad
m=1,2,3,\cdots\ ,
\label{jain-nu}
\ee 
can be mapped into a system of weakly interacting
``composite fermions'' at effective filling $\nu^*$,
\be
\frac{1}{\nu}\ \to\ \frac{1}{\nu^*}\ =\ \frac{1}{m} ,
\label{j-trans}
\ee 
by removing (or ``attaching'') $k$ quantum units of flux per particle
($k$ even). 
From (\ref{jain-nu}), the remaining effective magnetic field felt by
the composite fermions is:
\be
\B \ \to\ \B^* \ =\ B - \D \B\ ,\qquad \D \B = k\ 2\pi\rho_o \ .
\label{delta-b}
\ee
The relation between excluded magnetic field $\D \B$ and density is
the key point of Jain's argument.
The Lopez-Fradkin theory of the fractional Hall effect \cite{fradkin} 
implements this relation as
the equation of motion for the added Chern-Simons interaction, upon
tuning its coupling constant to $\k=1/k$, and taking the mean-field
approximation $\rho=\rho_o$.

Here we would like to stress that the Chern-Simon matrix model
provides another realization of the Jain composite-fermion transformation
(\ref{j-trans},\ref{delta-b}) for $m=1$.
For $k=0$, the matrix theory reduced to the eigenvalues $\l_a$
is equivalent to a system of free fermions in the lowest Landau level,
i.e. to $\nu^*=1$ \cite{karabali}\cite{cr}\cite{wiegmann}.
In the presence of the $\th$ background, the noncommutativity
of matrix coordinates
(\ref{nc-rel}) forces the electrons to acquire a finite area of order
$\th$, by the uncertainty principle, leading to the (semiclassical)
density $\rho_o=1/2\p\th$ (\ref{rho-def}) \cite{susskind}.
Using this formula of the density and the quantization of $\B\th$,
we re-obtain the Jain relation (\ref{delta-b}),
\be
\B\th \ = k\ \in\ \Z\ \ \to\ \ \B \ = \ k\ 2\pi\rho_o \ .
\label{delta-b-mcs}
\ee
Given that noncommutativity is expressed by the Gauss law of the matrix
theory, we understand that the mechanism for realizing
the Jain transformation is analogous to that of the Lopez-Fradkin theory, 
but it is expressed in terms of different variables.

The results of the Chern-Simons matrix theory were however limited, because
the (matrix analogues of) Jain states for $m=2,3,\dots$ could not be found. 
In the following, we shall find them in Maxwell-Chern-Simons matrix theory.


\subsection{General gauge-invariant states and their degeneracy} 

Consider first the case $k=1$.
The states in the lowest Landau level, i.e. the  
polynomials $\Phi_{\{n_1,\dots,n_N\}}\left(X,\psi\right)$ in
Eq. (\ref{HVR-states}), can be represented graphically as ``bushes'',
as shown in Fig.(1a).
The matrices $X_{ab}$ are depicted as oriented segments with indices at 
their ends and index summation amounts to joining segments
into lines, as customary in gauge theories.
The lines are the ``stems'' of the bush ending with one $\psi_a$, represented
by an open dot, and the epsilon tensor is the N-vertex
located at the root of the bush. 
Bushes have N stems  of different lengths:
$n_1 < n_2 <\cdots <n_N$. 
The position $i_\ell$ of one
$X$ on the $\ell$-th stem, $1\le i_\ell\le n_\ell$, 
is called the ``height'' of $X$ on the stem. 

\begin{figure}
\begin{center}
\includegraphics[width=10cm]{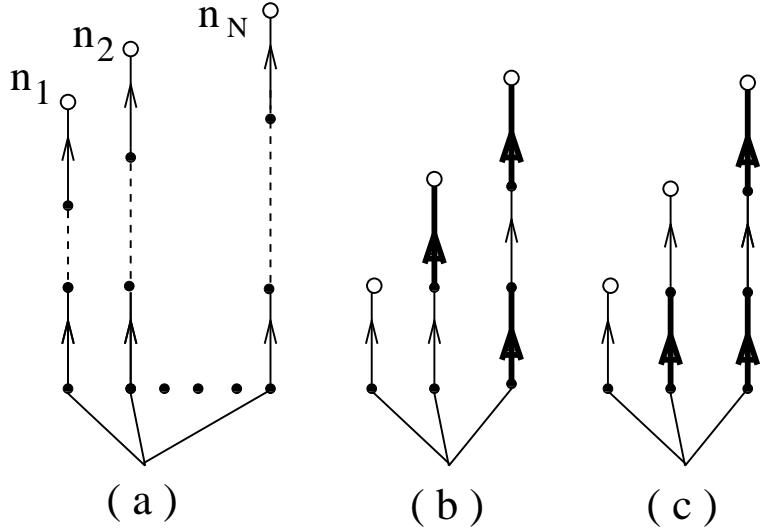}
\end{center}
\caption{Graphical representation of gauge invariant states: (a) 
general states in the lowest Landau level (cf. Eq.(\ref{HVR-states})); 
(b) and (c) examples in the second and third one for $N=3$.}
\label{fig1}
\end{figure}
 
The general solutions of the $k=1$ Gauss law (\ref{mcs-gauss},\ref{diff-gauss})
will be $\Phi$ polynomials involving both $X$ and $\ov{X}$: 
given that they transform in the same way under the gauge group
(cf. (\ref{diff-gauss}), the polynomials will again have the form of
bushes whose stems are  arbitrary words of $X$ and $\ov{X}$. 
Angular momentum and energy eigenstates are linear combinations of bushes with
given number ${\cal J}= N_X-N_{\ov{X}}$.
From the commutation relations (\ref{a-comm},\ref{b-comm}), 
energy and momentum eigenstates can be easily obtained by applying the
$A^\dagger_{ab}$ and $B^\dagger_{ab}$ operators (\ref{diff-op})
 to the empty ground state
$\Psi_o=\exp\left(-\Tr \ov{X}X/2 - \ov{\psi}\psi/2\right)$.
Their energy  $E=\B N_A$ 
and momentum ${\cal J}=N_B-N_A$ are expressed in terms of the number 
of $A^\dag$ and $B^\dag$ operators, $N_A$ and $N_B$ respectively.
The polynomial part $\Phi$ of the wave function is
thus expressed in the following variables:
\be
\Psi\ = \ e^{-\Tr\ \ov{X} X/2 -\ov{\psi} \psi/2}\ 
\Phi(\ov{B},\ov{A},\psi) \ ,\qquad
E=\B\ N_A\ , \quad J=N_B-N_A\ ,
\label{wf-def2}
\ee
 where
$\ov{B}=X-\de/\de\ov{X}$ and  $\ov{A}=\ov{X}-\de/\de X$ commute
among themselves, $[[\ov{A}_{ab},\ov{B}_{cd}]]=0$,
 and can be treated as $c$-number matrices.
Since their U(N) transformations are the same as those of $X,\ov{X}$,
they can be equivalently used to build the gauge invariant
bush states.
Examples of these general states  are drawn in Fig.(1b, 1c) for $N=3$: 
the variable $\ov{B}_{ab}$, replacing $X_{ab}$ 
in the lowest Landau level, is represented by a thin segment,
while $\ov{A}_{ab}$ is depicted in bold.
Upon expanding $\ov{A},\ov{B}$ in coordinates and derivatives acting
inside $\Phi$, one obtains in general a sum of $(X,\ov{X})$-bushes
as anticipated.

The form of the general $k=1$ gauge-invariant states suggests
a pseudo-fermionic Fock-space representation 
involving N ``gauge-invariant particles'', as it follows:
\begin{itemize}
\item
Each stem in the bush is considered as a ``one-particle state'' with quantum
numbers, $n_{Ai}, n_{Bi}$, characterizing individual energies and momenta
that are additive over the N particles, $N_A=\sum_{i=1}^N n_{Ai}$,
$N_B=\sum_{i=1}^N n_{Bi}$.
\item
Since two stems cannot be equal, one should build 
a Fermi sea of N such one-particle states.
\item
The one-particle states form again Landau levels with energies
$\eps_i=\B n_{Ai}$, but there are additional degeneracies at fixed
momentum with respect to the ordinary system; actually, 
in each stem, all possible
words of $\ov{A}$ and $\ov{B}$ of given length yield independent states, 
owing to matrix noncommutativity (assuming large values of N).
\end{itemize}

\begin{figure}
\begin{center}
\includegraphics[width=10cm]{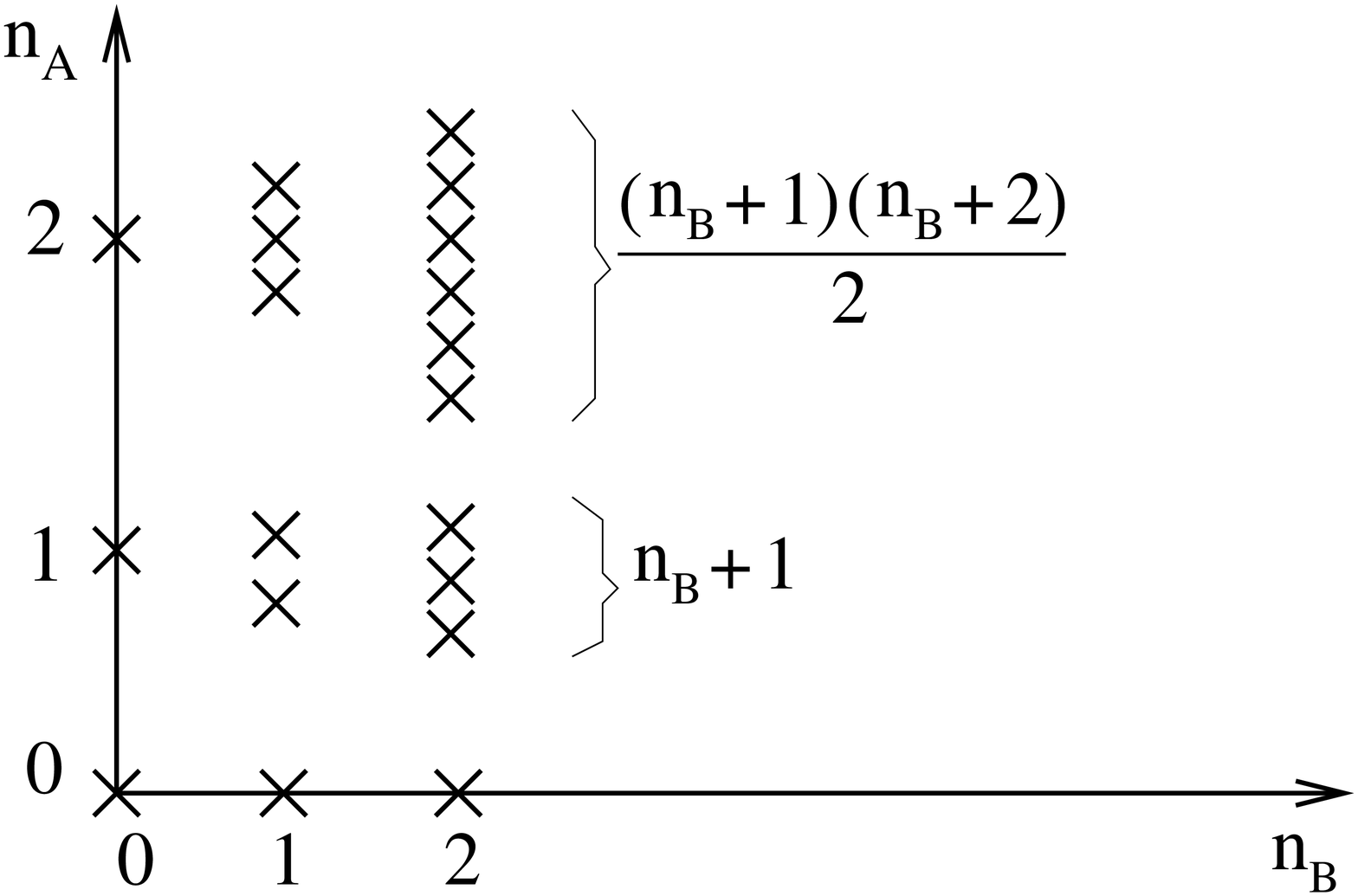}
\end{center}
\caption{Pseudo-fermionic Fock space representation of gauge invariant
states for $k=1$.}
\label{fig2}
\end{figure}

Such ``gauge invariant Landau levels'' 
are shown in Fig.(\ref{fig2}), together
with their degeneracies, $(n_A+n_B)!/n_A! n_B!$, given by the number words
of two letters with multiplicities $n_A$ and $n_B$.
These gauge invariant states should not be confused with
the Landau levels discussed in 
section two, that are relative to the states of the  $N^2$ gauge
variant ``particles'' with $X_{ab},\ov{X}_{ab}$ coordinates.
The analysis of some examples shows that the gauge invariant states
are many-body superpositions of the former $N^2$  states
that are neither bosonic nor fermionic and thus rather difficult to
picture. Instead, the interpretation in terms of $N$ fermionic
 ``gauge-invariant particles'' is rather simple and also convenient
for the physical limit $g=\infty$ of commuting matrices 
(to be discussed in section four).
Finally, the gauge invariant states solution of the  $k>1$ Gauss law are
given by tensoring $k$ copies of the structures just described, in complete
analogy with the lowest-level solutions (\ref{HVR-bose}).
Thus there are $k$  Fermi seas to be filled with N
``gauge-invariant particles'' each.

In the following, we are going to introduce a set of projections of the 
$g=0$ Maxwell-Chern-Simons theory that will reduce
the huge degeneracy of matrix states.

Degeneracies are better accounted for in a finite system, so we first 
modify the Hamiltonian to this effect.
For example, the quadratic confining potential $V_C$ (\ref{conf-pot})
permits degenerate states that have equal
energy and angular momentum -- this also occurs
in the ordinary Landau levels.
The problem can be solved by using finite-box boundary conditions,
that can be simulated by modifying the confining potential $V_C$
in the Hamiltonian (\ref{conf-pot}) as follows:
\be
V_C \ =\ \w\ \Tr\left(B^{\dag}\ B  \right)\ 
 +\ \w_n\ \Tr\left(B^{\dag n}\ B^n  \right)\ ,
\label{winf-bc}
\ee
for a given value of $n$.
The added operator $\Tr\left(B^{\dag n}\ B^n  \right)$
commutes with the $g=0$ Hamiltonian and
angular momentum and has the following spectrum:
when acting on stems, each $B_{ba}$ is a derivative that 
erases one $\ov{B}_{ab}$ matrix and fixes the indices at the loose ends
of the stem to $a$ and $b$ respectively.
Next, further $(n-1)$ derivatives act, with index summations, and finally
the length-$n$ strand  $\ov{B}^n_{ab}$ is added to complete
a new bush without cut strands.
On stems with $n_B \ge n$, this operator has a
diagonal action with eigenvalue $O(N^{n-1})$;
on other strands, it is non-diagonal with $O(1)$ coefficients.
Therefore, in the limit of large N and in the physical regime 
$n_A \ll n_B$, the confining potential (\ref{winf-bc}) effectively 
realizes the finite-box condition $n_{Bi} \le n$ for all Landau levels.

In a finite system of size $n$, the degeneracy of the $k$-th 
``gauge invariant Landau level'' 
is $O(n^k/k!)$ and the total degeneracy grows exponentially,
$O(\exp(n))$, for large energy. In presence of a quadratic confining
potential, it would grow exponentially with the energy and 
give rise to a Hagedorn transition at finite temperature.
Here one rediscovers a known property of matrix theories  
that makes them more similar to string theories  than to field theories of
ordinary matter \cite{mtheory}\cite{dzero}.

In our physical setting, we should consider this feature as a pathology of
the $g=0$ theory that should be cured in some
way. Actually, for $g>0$ the potential term in the Hamiltonian, 
$V=(g/8)\Tr \left[X,\ov{X}\right]^2$, tends to eliminate the
degeneracy due to matrix noncommutativity, as it follows. 
Consider a pair of degenerate states at $g=0$, that differ
for one matrix commutation, such those shown in Fig.(2b,2c),
and call their sum and difference $\Psi_+$ and $\Psi_-$, respectively.
For large values of $g$, the state $\Psi_+$ can have a finite energy,
while $\Psi_-$ will acquire a growing energy, corresponding
to the freezing of the degrees of freedom of the commutator.

Therefore, the Maxwell-Chern-Simons matrix theory for large
values of $g$ possesses degeneracies that are consistent
with ordinary two-dimensional matter; indeed, in section four
we shall show that the theory at $g=\infty$ 
reduces to the ordinary quantum Hall effect
with $O(1/r^2)$ interparticle interactions.
In conclusion, the matrix degeneracies at $g=0$ can be dealt with
by the theory itself by switching the $V$ potential on.

This fact is however not particularly useful 
from the practical point of view, because we do not presently know how
to compute the spectrum of the theory for $g>0$.
Precisely as in the original problem of the quantum Hall effect,
the free theory is highly degenerate and the degeneracy is broken
by interaction (potential).
The introduction of matrix variables would not appear as a great
improvement towards solving this problem, given that their degeneracies
are actually larger.

In spite of this, we shall find that matrix states do capture some
features of the quantum Hall dynamics.
In the following we shall introduce truncations
of the $g=0$ matrix theory that will eliminate most degeneracies
and will naturally select non-degenerate ground states that are 
in one-to-one relation with the Jain hierarchical states \cite{jain}.


\subsection{The Jain ground states by projection} 

We reconsider  the lowest Landau level condition (\ref{lll-cond})
$A_{ab} \Psi=0$, $\forall a,b$,
that singles out the Laughlin wave functions as the unique ground states
at filling fractions $\nu=1/(k+1)$.
Although apparently not gauge invariant, 
it follows from the gauge invariant condition of vanishing energy,
because the Hamiltonian, $H=2\sum_{ab}\ A^\dag_{ab}\ A_{ab}$, is
a sum of positive operators that should all individually vanish.

Consider now the weaker condition,
$\left( A_{ab} \right)^2 \Psi=0$, $\forall\ a,b$, 
allowing the $N^2$ gauge variant ``particles'' to populate
the first and second Landau level.
On the polynomial wave function, this projection reads: 
\be
\left(\frac{\de\ \ }{\de \ov{A}_{ab}}\right)^2
\ \Phi(\ov{A},\ov{B},\psi) \ =\ 0\ ,\qquad
\forall\ a,b \ .
\label{sll-cond}
\ee
The solutions are polynomials that are at most linear
in each gauge-variant component $\ov{A}_{ab}$:
one can think to an expansion of $\Phi$ in powers of $\ov{A}_{ab}$
that must stop at finite order as in the case 
of Grassmann variables.
The condition $\left( A_{ab} \right)^2 \Psi=0$ cannot be implemented
as a gauge invariant term in the Hamiltonian, because the corresponding
sum of positive operators, $\sum_{ab}\ A^{\dag 2}_{ab}\ A^2_{ab}$,
would not be gauge invariant.
Nevertheless, when acting on the bush states described in the
previous section (cf. Fig(\ref{fig1})), this condition
respects gauge invariance\footnote{ 
Instead, non-vanishing eigenstates of $A_{ab}^2$ are gauge variant.}.
In summary, this condition defines a gauge invariant truncation of
the $g=0$ matrix theory that cannot be easily realized by a change in
the Hamiltonian.

In the following we shall study the truncated
matrix theories that are defined by the 
projections: $\left( A_{ab} \right)^m \Psi=0$,
for $m$ taking the successive values $2,3,4,\dots$;
their wave functions contain the $N^2$ gauge variant
``particles'' filling the lowest $m$ Landau levels.

We first discuss the theory with second level projection $A^2=0$:
we outline the solutions of condition (\ref{sll-cond}) leaving the details
to appendix A.
Let us try to insert one or more $\ov{A}$ at points on 
the bush and represent them as bold segments, as in Fig.(\ref{fig1}).
The differential operator  (\ref{sll-cond})
acts by sequentially erasing pairs of bold lines on the bush in any order,
each time detaching
two branches and leaving four free extrema with indices fixed to 
either $a$ or $b$, with no summation on them.
For example, when acting on a pair of $\ov{A}$ located on the same stem, 
it yields a non-vanishing result: this limits to one $\ov{A}$ per stem.
Cancellations can occur for pairs of $\ov{A}$ on 
different stems, owing to the
antisymmetry of the epsilon tensor, as it follows:
\be
\left(A_a^b\right)^2\Phi =\cdots \ +\  \eps^{\dots i \dots j \dots}
\left(
\cdots M_{ia}\ N_{ja}\cdots V^b\ W^b \right)
\ +\ \cdots\ ,
\qquad (a,b\ {\rm fixed}),
\label{sll-canc}
\ee
that vanishes whenever $M=N$.
The analysis of appendix A shows that there cannot be
further cancellations involving linear combinations of different
bushes. 
Therefore, the general solution of (\ref{sll-cond}) is a bush involving one
$\ov{A}$ per stem (max N matrices in total), all of
them located at the same height on the stems, as follows:
\ba
\Phi^{(II)}_{\{n_1,\dots,n_\ell; p; n_{\ell+1},\dots,n_M\}} 
&=& \eps^{i_1\dots i_N} \ 
\prod_{k=1}^{\ell}\ 
\left(\ov{B}^{n_k}\psi\right)_{i_k} \
 \prod_{k=\ell+1}^{N}\ 
\left(\ov{B}^p\ov{A}\ \ov{B}^{n_k} \psi\right)_{i_k}\ , 
\nl
&&\quad 0\le n_1 < \cdots < n_\ell\ ,\quad
0\le n_{\ell+1} < \cdots < n_N\ . 
\label{two-ll}
\ea

These states can be related to Slater determinants
of the ordinary Landau levels: 
assuming diagonal expressions for both $\ov{B}$ and $\ov{A}$,
the matrix states become Slater determinants of N electron
one-particle states \cite{jain}\cite{numeric}.
This relation is surjective in general, because
states differing by matrix orderings get identified;
however, for states of form of Eq. (\ref{two-ll}),
the matrix degeneracy is limited to the $p$ dependence.
This shows how the projection $A^2=0$ works in reducing
degeneracies.

Let us analyze the possible matrix states in the $A^2=0$ theory 
with finite-box conditions, referring to Figs.(\ref{fig1}, \ref{fig2}) 
for examples.
The most compact state corresponds to {\it homogeneous} filling 
all the allowed states in the first and second Landau levels with $N/2$
``gauge invariant particles'' each; it reads:
\be
\Phi_{1/2, \ gs} = \eps^{i_1\dots i_N} \ 
\prod_{k=1}^{N/2}\ 
\left(\ov{B}^{k-1}\psi\right)_{i_k} \ 
\prod_{k=1}^{N/2}\ 
\left(\ov{A}\ \ov{B}^{k-1} \psi\right)_{i_{N/2+k}}\ , 
\label{two-jain}
\ee
with angular momentum ${\cal J}=N(N-4)/4$.
One easily sees that this state is non-degenerate for boundary conditions
enforcing maximal packing, $n_{Bi}\le N/2$, due to the vanishing of the $p$ 
parameter in (\ref{two-ll}).
Assuming homogeneity of its density, we can assign it the filling fraction
$\nu^*=2$  using (\ref{nu-def}). 

Let us now discuss the states in the $A^2=0$ theory for generic $k$ values.
Gauge invariant states should be products of $k$
bushes, as in (\ref{HVR-bose}): they survive the 
projection (\ref{sll-cond}),  provided that the two derivatives 
always vanish when distributed over all bushes.
Given the product state with one bush of type (\ref{two-jain}),  
obeying $A^2\ \Phi_{1/2, \ gs}=0$,
\be
\Phi_{k+1/2, \ gs} \ = \ \Phi_{k-1, \ gs} \ \Phi_{1/2, \ gs} \ ,
\label{two-k-jain}
\ee
the other factor involving $k-1$ bushes should
satisfy $A\ \Phi_{k-1,\ gs} =0$ and actually be the Laughlin state
(\ref{HVR-states}).
The state (\ref{two-k-jain}) is also non-degenerate with appropriate 
tuning of the boundary potential.
From the ${\cal J}$ value, one can assign  the 
filling fraction\footnote{
Keeping in mind the contribution of $1$ from the Vandermonde of
the integration measure.},
$1/\nu=k+1/2$, to this state.

We thus find the important result that the projected
Maxwell-Chern-Simons theory possesses non-degenerate ground states 
that are the matrix analogues of the Jain states  obtained by 
composite-fermion transformation at $\nu^*=2$, 
Eqs. (\ref{jain-nu},\ref{j-trans}).
The matrix states (\ref{two-k-jain},\ref{two-jain}) 
would actually be equal to Jain's wave functions, if the
$\ov{A},\ov{B}$ matrices were diagonal: the $\psi$ dependence
would factorize and the matrix states 
reduce to the Slater determinants of Jain's wave functions
(before their projection to the lowest Landau level) 
\cite{jain}\cite{numeric}.
Indeed, the diagonal limit can be obtained as follows.
We note that the derivatives present in the expressions (\ref{wf-def2}) of 
$\ov{A}$ and $ \ov{B}$ vanish when acting on the states (\ref{two-jain})
due to antisymmetry of the epsilon tensor\footnote{
Using the graphical rules introduced before
(cf. Fig. \ref{fig1}), this simplification is found for all states
where Landau levels are homogeneously (i.e. completely) filled 
with non-increasing number of particles, namely 
$N_1\ge N_2 \ge\cdots \ge N_m$, with $\sum_{i=1}^m N_i=N$.}:
in the expression of these states we can replace, 
$\ov{B}\to X,\ov{A}\to \ov{X}$.
Therefore, the Jain and matrix states 
become identical in the limit of diagonal $X,\ov{X}$, that
is realized for $g\to\infty$ as discussed in section four.

The correspondence extends to the whole Jain series:
the other $\nu^*=m$ non-degenerate ground states are respectively
obtained in the theories with $A^m=0$ projections.
Before discussing the generalization,
let us analyze the other allowed states by the $A^2=0$ projection.
They are obtained by relaxing the boundary conditions
for (\ref{two-jain}), i.e. by reducing the density of the system, allowing
for lower fillings of the ``gauge invariant Fermi sea''.
The non-degenerate Laughlin ground state and its quasi-hole 
are clearly allowed states in the lowest level (cf. section 3.1).
The quasi-particle over the Laughlin state is 
obtained by having one particle in the second Landau level,
leading to the form (\ref{two-ll}) involving 
one $\ov{A}$ only, i.e. $\ell=N-1, p=0, n_N=0$,
\ba
\Phi_{k,\ 1qp} &=& \Phi_{k-1,\ gs}\ 
\Phi^{(II)}_{1, \ 1qp}\ ,
\nl
\Phi^{(II)}_{1, \ 1qp} &=& \eps^{i_1\dots i_N} \ 
\left(\ov{A}\psi\right)_{i_N} \ 
\prod_{k=1}^{N-1}\ \left(\ov{B}^{k-1}\psi\right)_{i_k} \ .
\label{lau-qp}
\ea
This is a quasi-particle in the inner part of the Laughlin fluid,
it is non-degenerate and has the gap $\D E_{1qp} = \B$ 
(disregarding the confining potential) and $\D {\cal J} =-N$.
Other quasi-particles are density rings that can be
degenerate due to the free $p$ parameter in (\ref{two-ll}).
Multi quasi-particle states are obtained by inserting more than
one $\ov{A}$ in $\Phi^{(II)}$, on different stems of the bush, according
to (\ref{two-ll}):
$\Phi_{k,\ \ell qp} = \Phi_{k-1,\ gs}\ \Phi^{(II)}_{1, \ \ell qp}$.
Their energy is linear in the number of quasi-particles.
We thus find that the projected $g=0$ Maxwell-Chern-Simons matrix theory 
reproduces the Jain composite-fermion correspondence 
also for quasi-particle excitations \cite{jain}, but with additional
degeneracies.

Let us not proceed to find the states in the $g=0$ theory with
higher projections. 
In the $A^3=0$ theory, the $k=1$ bushes may have two $\ov{A}$ matrices
per stem at most, obeying the following rules 
(proofs are given in Appendix A):
\begin{itemize}
\item
If the bush has only one $\ov{A}$ per stem, i.e. for second-level
fillings, the $\ov{A}$'s can stay on the stems at two values 
of the height, i.e. can form two bands.
\item
If there are stems with both one and two $\ov{A}$'s, then 
the $\ov{A}$'s can form two bands, with the extra condition for
single-$\ov{A}$ stems that their $\ov{A}$'s should stay on the lowest band.
\end{itemize}
The first rule implies that 
the earlier $\nu=2^*$ homogeneous state (\ref{two-jain})
becomes degenerate in the $A^3=0$ theory at the same density.
On the other hand, the $A^3=0$ theory admits a
maximal density state with $N/3$ gauge-invariant
particles per level, that is unique due to the second rule:
\be
\Phi_{1/3, \ gs} = \eps^{i_1\dots i_N} \ 
\prod_{k=1}^{N/3}\ \left[
\left(\ov{B}^{k-1}\psi\right)_{i_k} \ 
\left(\ov{A}\ \ov{B}^{k-1} \psi\right)_{i_{k+N/3}}
\left(\ov{A}^2 \ \ov{B}^{k-1} \psi\right)_{i_{k+2N/3}}
\right]\ .
\label{three-jain}
\ee
This state corresponds to filling fraction $\nu^*=3$.
Next, the product states,
\be
\Phi_{k+1/3, \ gs} \ = \ \Phi_{k-1, \ gs} \ \Phi_{1/3, \ gs} \ ,
\label{m-3-jain}
\ee
obeys the $A^3=0$ condition for $k>1$: these ground states
realize the Jain composite-fermion construction for $\nu^*=3$ and have
the expected filling fraction $\nu=m/(mk+1)$ for $m=3$.

The pattern repeats itself in the $A^4=0$ theory (cf. appendix A): 
there are three
$\ov{A}$'s per stem at most, that can form up to three bands; however, if
single and/or double-$\ov{A}$ stems are present together with the 
three-$\ov{A}$ stems, the $\ov{A}$'s of the former stems
should stay on the lowest bands.
Therefore, the maximal density state is again unique, having form analogous
to (\ref{three-jain}) and filling $\nu^*=4$. 

In conclusion, the $A^m=0$ projected theory possesses the following
non-degenerate ground states with Jain fillings $\nu=m/(mk+1)$:
\be
\Phi_{k+1/m, \ gs}\ =\  \Phi_{k-1, \ gs} \ \Phi_{1/m, \ gs} \ ,
\label{m-k-jain}
\ee
where
\be
\Phi_{1/m, \ gs} =  \eps^{i_1\dots i_N} \ 
\prod_{k=1}^{N/m}\ \left[
\left(\ov{B}^{k-1}\psi\right)_{i_k} \ 
\left(\ov{A}\ \ov{B}^{k-1} \psi\right)_{i_{k+N/m}}
\cdots
\left(\ov{A}^m \ \ov{B}^{k-1} \psi\right)_{i_{k+(m-1)N/m}}
\right]\ .
\label{m-jain}
\ee
In the $A^m=0$ theory, the
lower density states that were non-degenerate in the 
$A^k=0$ theories, $k<m$, become degenerate.
Nevertheless, there are non-degenerate quasi-particles
of the $(m-1)$ Jain state just below.

In conclusion, we have found that the ground states with homogeneous
fillings of the properly projected Maxwell-Chern-Simons matrix theory 
reproduce the Jain pattern of the composite fermion transformation.
These matrix states are unique solutions for certain (maximal) values
of the density, while Jain states are 
judiciously chosen ansatzs among many possible multi-particle states
of the ordinary Landau levels.

These results indicate that the Jain composite-fermion
excitations have some relations with the D0-brane degrees of freedom
and their underlying gauge invariance. 
Both of them have been described as dipoles. According to Jain
\cite{jain} and Haldane-Pasquier \cite{pasquier}, 
the composite fermion can be considered as the bound state
of an electron and a hole (a vortex of the electron fluid): the reduced
effective charge would then account for the smaller effective magnetic field
$\B^*$  (\ref{delta-b}) felt by these excitations.
On the other side, matrix gauge theories, such as the Maxwell-Chern-Simons
theory, are equivalent to noncommutative
theories whose fundamental degrees of freedom are dipoles.
Clearly, a better understanding of the potential term $\Tr[X,\ov{X}]^2$
in our matrix theory is necessary to clarify the dipole description.

We finally remark that the matrix coordinates are less
noncommutative on the Jain states then on the Laughlin ones.
Indeed, the general form of the Gauss law (\ref{mcs-gauss})
can be rewritten in terms of $X,\ov{X},A,\ov{A}$ as follows:
\be
\left[X,\ov{X}\right] \ +\ \frac{2}{B}
\left[\ov{X},A\right] \ +\  \frac{2}{B} \left[\ov{A},X\right] \ = \
2\left( \theta - \frac{1}{B}\psi\otimes\ov{\psi} \right)\ . 
\label{not-non-com}
\ee
On the Laughlin states belonging to the lowest Landau level,
this reduces to the coordinates noncommutativity (\ref{nc-rel}),
because $A=\ov{A}=0$; on states populating higher levels, there are
other terms contributing to noncommutativity besides the matrix 
coordinates. In section four, we shall discuss the theory in the
opposite $g=\infty$ limit, where $[X,\ov{X}]=0$, and thus 
non-commutativity is entirely realized between coordinates and momenta.


\subsection{Generalized Jain's hierarchical states}

In the $A^m=0$ projected theories with $m\ge 3$, there are other
solutions of the Gauss law for $k>1$ besides the Jain states (\ref{m-k-jain}).
Any combination of the $k=1$ solutions (\ref{m-jain}) 
is possible, as follows:
\ba
\Phi_{\frac{1}{p_1}+\cdots+\frac{1}{p_k}, \ gs} 
&=& \prod_{i=1}^k\ \Phi_{\frac{1}{p_i}, \ gs}\ ,
\nl
\frac{1}{\nu} &=&  1\ + \ \sum_{i=1}^k\ \frac{1}{p_i} \ .
\label{ext-jain}
\ea
In this equation, we also wrote the associated filling fractions
using Eq.(\ref{nu-def}), i.e. assuming homogeneous densities.
The states (\ref{ext-jain}) obey the condition $A^q=0$
with $q=1+\sum_{i=1}^k (p_i-1)$.  
The Jain mapping to a single set of  $\nu^*=q$
effective Landau levels does not hold for these generalized states.
Actually, analogous states were considered by Jain as well \cite{jain}, 
and disregarded
as unlikely further iterations of the composite-fermion transformation. 
In the matrix theory, we seek for arguments to disregard them as well.

Let us compare the generalized (\ref{ext-jain}) and 
standard (\ref{m-k-jain}) Jain states at fixed values of the background $k$
(keeping in mind that the physical values are $k=2,4$). 
The energy of the generalized states is additive in the 
$\nu^*=p_i$, $k=1$, blocks and reads:
\be
E_{\frac{1}{p_1}+\cdots+\frac{1}{p_k}, \ gs} 
= \frac{\B N}{2} \ \sum_{i=1}^k\ (p_i-1)\ +\ V_C\ .
\label{en-gen}
\ee
The analysis of some examples of fillings and energies  makes it clear that
these additional solutions have in general higher energies for
the same filling or are more compact for the same energy
than the standard Jain states (\ref{m-k-jain}) (see Table \ref{jain-tab}).
States of higher energies  are clearly irrelevant at low temperatures.
Furthermore, higher-density states strongly deviate from the
semiclassical incompressible fluid value $\nu=1/(k+1)$ for background 
$\B\th=k$, that is specific of the Laughlin factors \cite{susskind}.
This fact indicates that they might not be
incompressible fluids with uniform densities.
Further discussion of this point is postponed to section four.

\begin{table}
\begin{center}
\[
\begin{array}{c|lccc|l|l}
m  & \# p_i=1 & \# p_i=2 & \# p_i=3 & \# p_i=4 & 1/\nu & E/\B \\
\hline 
1  & k     &      &       &      &  k+1      &  0  \\
\hline 
2  & k-1   & 1    &       &      &  k+ 1/2   &  N/2   \\
\hline 
3  & k-2   & 2    &       &      &  k        &  N   \\
3  & k-1   & 0    & 1     &      &  k+1/3    &  N  \\
\hline 
4  & k-3   & 3    &       &      &  k-1/2    &  3N/2  \\
4  & k-2   & 1    & 1     &      &  k-1/6    & 3N/2   \\
4  & k-1   & 0    & 0     & 1    &  k+1/4    & 3N/2   
\end{array}
\]
\end{center}
\caption{Examples of generalized  (\ref{ext-jain}) and standard 
(\ref{m-k-jain}) Jain states for fixed value of
$k$, ordered by Landau level $m$ with corresponding fillings 
$\nu$ and energies $E$ (disregarding the confining potential). 
Note that the experimentally  relevant values are $k=2,4$ \cite{jain}. }
\label{jain-tab}
\end{table}


\section{$g\to\infty$ limit and electron theory}

In this section we switch on the potential 
$V=-(g/2) \Tr [X_1,X_2]^2$ in the Hamiltonian (\ref{mcs-ham}) 
and perform the $g\to\infty$ limit.
The potential is a quartic interaction between the
matrices that does not commute with the Landau term,
$\B \Tr\left( A^\dag A\right) $: thus,
the $g=0$ eigenstates obtained in the previous section
by filling a given number of Landau levels will evolve
for $g>0$ into mixtures of states.

At the classical level, the $V$ potential suppresses the matrix
degrees of freedom different from the eigenvalues, and
projects them out for $g\to\infty$.
This can be seen by using the Ginibre decomposition of complex matrices 
\cite{ginibre}, which reads:
$X=\ov{U}(\L+R)U$, where $U$ is unitary (the gauge degrees of 
freedom), $\L$ diagonal (the eigenvalues) and $R$ complex upper triangular
(the additional d.o.f.).
Inserting this decomposition in the potential, we find for $N=2$:
\be
V=\frac{g}{8}\Tr\left[X,\ov{X}\right]^2 = 
\frac{g}{4}\vert r\vert^4 + 
\frac{g}{4}\vert\ov{r}\left(\l_1-\l_2\right)\vert^2\ ,
\qquad
X=\left(\begin{array}{cc} \l_1& r\\ 0 & \l_2 \end{array}\right)\ .
\label{v-form}
\ee
Thus for large $g$, the variable $r$ is suppressed.
For general $N$, the potential 
kills all the $N(N-1)$ real degrees of freedom contained in the $R$ matrix.

Let us now discuss the matrix theory in the
$g=\infty$ limit, i.e. for $R=0$: $X$ and $\ov{X}$ commute among themselves 
(they are called ``normal'' matrices \cite{wiegmann}) and can be 
diagonalized by the same unitary transformation:
\ba
&& X=\ov{U}\L U\ ,\quad \ov{X}=\ov{U}\ov{\L} U \ , \qquad 
\L = diag\ \left(\l_a\right)\ ,
\nl
&&\left[X,  \ov{X}\right] = 0\ .
\label{n-mat}
\ea
In the $g=\infty$ limit, we analyze the theory following a different strategy
from that of section 3: we fix the gauge invariance, solve the Gauss law 
at the classical level and then quantize the remaining variables, 
which are the complex eigenvalues $\l_a$ and their conjugate momenta
$p_a$, following the analysis of Refs. \cite{poly97}\cite{park}.
We take the diagonal gauge for the matrix coordinates
and decompose the momenta $\Pi,\ov{\Pi}$, 
in diagonal and off-diagonal matrices, respectively called $p$  and $\G$:
\be
X=\L \ , \qquad \P=p+\G\ ,\quad \ov{\P}=\ov{p}+\ov{\G}\ .
\label{l-expr}
\ee
The Gauss law constraint (\ref{mcs-gauss}) can be rewritten:
\ba
\left[X,\P \right]+ \left[\ov{X},\ov{\P} \right] &=& 
-i\ \B\th + i\ \psi\otimes\ov{\psi}\ ,
\nl
\left(\l_a-\l_b\right) \G_{ab} + 
\left(\ov{\l}_a-\ov{\l}_b\right) \ov{\G}_{ab}
&=& -i \left(k\ \d_{ab}- \psi_a\ \ov{\psi}_b\right)\ .
\label{eigen-gauss}
\ea
The second of (\ref{eigen-gauss}) implies 
$\vert \psi_a \vert^2=k$ for any value of $a=b$.
We can further fix the remaining $U(1)^N$ gauge freedom  by
choosing $\psi_a=\sqrt{k}$, $\forall a$, such that the r.h.s.
of Eq. (\ref{eigen-gauss}) becomes proportional to $(1-\d_{ab})$.

Therefore the Gauss law completely determines the off-diagonal momenta:
their rotation invariant form is,
\be
\G_{ab}\ =\ \frac{i k}{2}\ 
\frac{\ov{\l}_a-\ov{\l}_b}{\vert \l_a-\l_b\vert^2}\ ,
\qquad\quad a\neq b\ .
\label{gamma-val}
\ee
By inserting this back into the Hamiltonian (\ref{mcs-ham}), we find that
diagonal and off-diagonal terms decouple and we obtain,
\ba
H &=& 2\ \Tr \left[\left(\frac{\ov{X}}{2} -i\ \P\right)
\left(\frac{X}{2} +i\ \ov{\P}\right) \right] \ 
\nl
 &=& 2\ \sum_{a=1}^N \ \left(\frac{\ov{\l}_a}{2} -ip_a\right)
\left(\frac{\l_a}{2} +i\ov{p}_a\right) \ +\ 
\frac{k^2}{2}\ \sum_{a \neq b =1}^N\ 
\frac{1}{\vert \l_a-\l_b\vert^2}\ .
\label{inf-ham}
\ea
The same result is obtained starting from the Lagrangian (\ref{mcs-action})
and solving for $A_0$ in the gauge $X=\L$ at $g=\infty$ \cite{park}.

Therefore, the theory reduced to the eigenvalues corresponds
to the ordinary Landau problem for N electrons plus an induced
two-dimensional Calogero interaction.
Note also that the matrix measure of integration
(\ref{int-meas}) reduces to the ordinary expression
after incorporating one Vandermonde factor $\D(\l)$ in the wave
functions \cite{wiegmann}.
The occurrence of the Calogero interaction 
is a rather common feature of matrix theories reduced to 
eigenvalues: the induced interaction is analog to the 
centrifugal potential appearing in the radial Schroedinger equation.
In the present case, the interaction is two-dimensional, owing
to the presence of two Hermitean matrices, and thus it is
rather different from the exactly solvable one-dimensional case 
\cite{poly1}\cite{poly5}.

We conclude that the Maxwell-Chern-Simons matrix theory in the
$g=\infty$ limit makes contact with the physical problem 
of the fractional quantum Hall effect: the only difference
is that the Coulomb repulsion $e^2/r$  
is replaced by the Calogero interaction $k^2/r^2$.
Numerical results \cite{laugh}\cite{haldane} \cite{jain}\cite{numeric}
indicate that quantum Hall incompressible
fluid states are rather independent of the detailed form of the
repulsive potential at short distance, for large $\B$.
In particular, the Calogero
potential does not have the long-range tail of the Coulomb interaction and
is closer to the class of much-used Haldane short-range potentials 
\cite{haldane}. Although the physics of incompressible fluids is universal,
the form of the potential might affect the detailed quantitative
predictions of the theory for some quantities such as the gap:
this issue is postponed to the future.

Some remarks are in order:
\begin{itemize}
\item
The physical condition imposed by the Gauss law (\ref{eigen-gauss})
is still that outlined in
section 3.2: it forces the electrons to stay apart by
locking their density to the value of the background parameter $k$.
The solution of this constraint is however rather different at 
the two points $g=0$ and $g=\infty$:
for $g=0$, it is the geometric, or kinematic, condition of noncommutativity
(\ref{nc-rel}),
while at $g=\infty$ this is a dynamical condition set by a repulsive
potential with appropriate strength. 
\item
Such dynamical condition is far more complicate to solve, and it 
allows many more excited states than the kinematic condition;
there are many more available states  in the 
lowest Landau level at $g=\infty$ than in the $g=0$ matrix theory.
\item
Note also that the $g=\infty$ theory is not, by itself, less
difficult than the ab-initio quantum Hall problem: the gap is
non-perturbative and there are no small parameters.
The advantage of embedding the problem into the matrix theory is 
that of making contact with the solvable $g=0$ limit, as we discuss in
the next section. 
\end{itemize}

\begin{figure}
\begin{center}
\includegraphics[width=10cm]{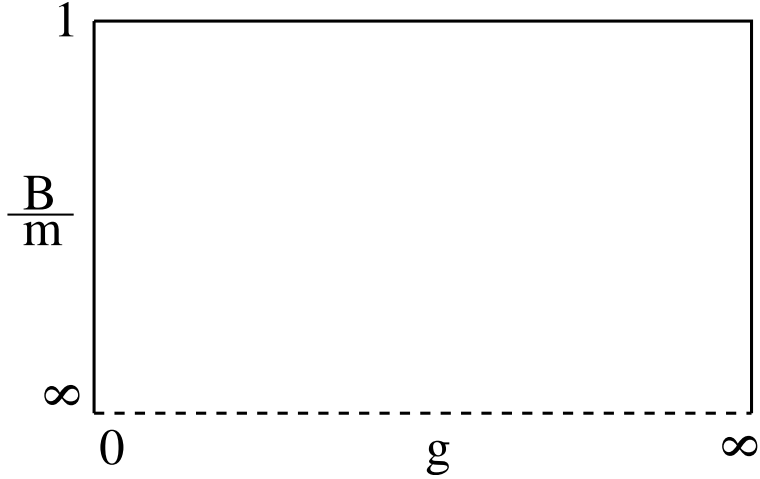}
\end{center}
\caption{Phase diagram of the Maxwell-Chern-Simons matrix theory.
The axes $g=0$ and $g=\infty$ have been discussed in sections
3 and 4, respectively. The Chern-Simons matrix model sits at the 
left down corner.}
\label{fig3}
\end{figure}


\section{Conjecture on the phase diagram and conclusions}

In Figure (\ref{fig3}) we illustrate the phase diagram of the
Maxwell-Chern-Simons matrix theory as a function of its
parameters $\B/m$ and $g$.
The quantized background charge $\B\th=k$ is held fixed 
over the diagram together with the parameters $\w,\w_n$ in
the confining potential (\ref{winf-bc}).
 
The axes $g=0$ and $g=\infty$ have been discussed in sections
3 and 4, respectively. For $g=0$, the theory is 
solvable and displays a set of states that are in 
one-to-one relation with the Laughlin and Jain ground states
with filling fractions $\nu=m/(mk+1)$.
These non-degenerate states  can be selected by choosing
the appropriate projection $A^m=0$ and the value of 
$k$, and by tuning $\w,\w_n$. 
For $g=\infty$, we found that the theory describes the real
fractional Hall effect, but we do not know how to solve
the Calogero interaction and find the ground states.

Let us consider the evolution of one Jain state as $g$ is switched on,
while keeping the other parameters fixed.
Given that the potential $\Tr\left[X,\ov{X}\right]^2$
does not commute with the $g=0$ Landau Hamiltonian,
this state will mix with other ones.
If it remains non-degenerate as $g$ grows up
to infinity, we can say that the matrix theory remains in 
the same universality class and that the qualitative features found at
$g=0$ remain valid in the physical limit  $g=\infty$.
In the case of level crossing at some finite
value $g=g^*$, the two regimes of the theory are unrelated.

Unfortunately, we do not presently have a method of solution 
of the $g\neq 0$ Hamiltonian, even approximate, 
that could prove the non-degeneracy of Jain states for $g>0$ 
and establish the physical 
relevance of the Maxwell-Chern-Simons matrix theory.

Nevertheless, we would like to conjecture that the Laughlin and Jain
states at $g=0$ do remain non-degenerate. 
Namely, that there is no phase transition
at finite $g$ values when the theory is tuned  on such
ground states at $g\sim 0$ (by appropriate choices of $m,k,\w,\w_n$).

Our conjecture is indirectly supported by the numerical results 
by Jain and others \cite{haldane}\cite{jain} \cite{numeric}, 
through the following classical argument.
These authors found that the Laughlin and Jain states
in the quantum Hall effect are very close to the exact numerical
ground states for a variety of short-range potentials,
including the Calogero one realized at $g=\infty$.
Now, consider the $g>0$ evolution of the Jain
matrix ground states:
the effect of the potential can be seen, at the
classical level, as that of eliminating the additional matrix
d.o.f. and make both $X,\ov{X}$ matrices  diagonal 
(up to a gauge transformation, cf. section 4).
In this case, the Jain matrix states become Slater determinants 
of Hall states (cf. section 3.4) and exactly
reduce to the expressions introduced by Jain \cite{jain}.
Therefore, it is rather reasonable to expect that the evolution
the $g=0$ matrix states will bring them into the diagonalized, i.e.
original Jain states at $g=\infty$, up to small deformations.

On the contrary, other states such as those of the generalized 
Jain hierarchy (cf. section 3.5), 
that have no counterpart in the $g=\infty$ theory,
are likely to become degenerate at finite $g$.

In conclusion, our conjecture of smooth evolution of matrix Jain states
is supported
by the numerical analyses of the Jain composite-fermion theory. 
Let us add some comments:
\begin{itemize}
\item
The projected $A^m=0$ theory describes the $\nu^*=m$ Jain matrix state but
does not allow to discuss its gap, because its  quasi-particles live in 
the theory with higher projection, $A^{m+1}$, where the $\nu^*=m$ state 
is itself degenerate.
The projections complicate the $g=0 \leftrightarrow\infty$ correspondence
and might be relaxed at some point of the $g$-evolution; this
remains to be understood.
\item
As discussed in section 3, the limit $\B\to\infty$
cannot be taken at $g=0$, because quasi-particle excitations and
Jain states in the matrix theory have energies of $O(\B)$ and
would be projected out.  Instead, the limit $\B=\infty$ can surely
be taken in the $g=\infty$ physical theory
(holding $k=\B\th$ fixed), because the fractional
quantum Hall states are known to remain stable.
This implies that the two limits are ordered: the correct sequence is
$\lim_{\B\to\infty}\ \lim_{g\to\infty}\Psi$, and the opposite choice 
is cut out in the phase diagram of Fig.\ref{fig3}.
\item
The projection to the lowest Landau level, $\B \to\infty$ at
$g\gg 1$ should also
transform the Jain states into incompressible fluids with constant
densities corresponding to the filling fractions assigned before.
It is rather clear that the Jain matrix states at $g=0$ (section 3.4)
are not uniform: they are multiple-droplet
states similar to those considered in \cite{poly3}.
On the other hand, the matrix Laughlin states (\ref{HVR-bose}) do
correspond to incompressible
fluids in the lowest Landau level as shown in the Refs.
\cite{susskind}\cite{poly1}.
\end{itemize}

In summary, in this paper we have generalized the 
Susskind-Polychronakos proposal of noncommutative Chern-Simons
theory and matrix models. We have found:
\begin{itemize}
\item A description of the expected Jain states and their
quasi-particle excitations within a matrix generalization of the Landau 
levels.
\item An interesting phase diagram, parametrized by the additional coupling
$g$, with a manifestly physical limit for the matrix theory
at $g=\infty$.
\end{itemize}
Reliable methods of solution for the potential 
$g\ \Tr\left[X,\ov{X}\right]^2$ are needed to understand the
phase diagram, verify
the proposed physical picture and allow for the physical applications.
Recent developments in the analysis of multi-matrix theories 
\cite{agarwal} may provide new tools for tackling this problem.

\bigskip

{\large \bf Acknowledgments}

We warmly thank A. Polychronakos for many discussions and  
suggestions on this work. 
We also thank M. Riccardi, D. Seminara and G. R. Zemba
for interesting discussions.
A. Cappelli thanks the KITP, Santa Barbara, for hospitality.
This research was supported in part by the National Science  
Foundation under Grant No. PHY99-07949.
I. D. Rodriguez thanks the EC program Alban of Ph-D scholarships for
Latin American students.
This work was partially funded by the EC Network contract 
HPRN-CT-2002-00325, 
``EUCLID: Integrable models and applications: 
from strings to condensed matter'', and by the ESF programme 
``INSTANS: Interdisciplinary Statistical and Field Theory Approaches 
to Nanophysics and Low Dimensional Systems''.

\appendix


\section{Projections of matrix Landau states} 

In this appendix we prove some properties of the projections of states
described in section 3.4.
Let us first show that the general solution of the second Landau
level projection (\ref{sll-cond}),
\be
\left(\frac{\de\ \ }{\de \ov{A}_{ab}}\right)^2
\ \Phi(\ov{B},\ov{A},\psi) \ =\ 0\ ,\qquad
\forall\ a,b \ .
\label{sll-app}
\ee
and of the $k=1$  Gauss law (\ref{diff-gauss}) is given by the expression
(\ref{two-ll}).
We start from the gauge invariant expressions involving N  fields $\psi$,
\be
\Phi= \eps^{a_1\dots a_N} \ 
\left(M_1\psi\right)_{a_1}\cdots \left(M_N\psi\right)_{a_N}\ ,
\label{eps-app}
\ee
where the $M_i$ are polynomials of $\ov{B}$ and $\ov{A}$.
In this appendix, we repeatedly use the graphical description of
these expressions in terms of bushes as shown in Fig.\ref{fig1}.
Upon expanding (\ref{eps-app}) into monomials, we get a sum of bushes:
\be
\Phi= \eps^{a_1\dots a_N}
\left(P_1\psi\right)_{a_1}\cdots \left(P_N\psi\right)_{a_N} +\ 
b\ \eps^{a_1\dots a_N}
\left(Q_1\psi\right)_{a_1}\cdots \left(Q_N\psi\right)_{a_N} +\ 
c\ \eps\dots R_1\cdots R_N +\ \cdots ,
\label{sum-app}
\ee
where the monomials in a bush, e.g. the $\{P_i \}$, are all
different among themselves, and two sets of monomials,
e.g.  $\{P_i \}$ and  $\{Q_j \}$,  
differ in one monomial (stem) at least.

The two derivatives in (\ref{sll-app}) act in all possible ways on
the stems of the bushes, and can be represented by primed expressions,
i.e. $P_i',P_i'',\dots$. 
Let us momentarily take bushes made of two stems, i.e. $N=2$:
\ba
\Phi''
&=& \eps_{ab}\left(P_1''\psi_a\ P_2\psi_b + 2 P_1'\psi_a\ P_2'\psi_b + 
P_1\psi_a\ P_2''\psi_b \right) 
\nl
&+& b\ \eps_{ab}\left(Q_1''\psi_a\ Q_2\psi_b + 2 Q_1'\psi_a\ Q_2'\psi_b + 
Q_1\psi_a\ Q_2''\psi_b \right) +\cdots .
\label{der-app}
\ea
We check the possibility of cancellations 
between terms belonging to two different bushes: these cannot occur
between terms with the same pattern of derivatives, i.e. $P_1\ P_2''$ and 
$Q_1\ Q_2''$, because at least one monomial is different between the two
bushes: $P_1\neq Q_1$ or $P_2\neq Q_2$.
There can be cancellations between terms that have different derivatives,
i.e. $P_1''\ P_2 + b\ Q_1'\ Q_2'=0$, but then the symmetric
term would not cancel, $P_1\ P_2'' + b\ Q_1'\ Q_2' \neq 0$.
We conclude that there cannot be complete cancellations between two
bushes and that each bush should vanish independently.

Consider now the action of derivatives on the stems of a single bush;
the terms with two derivatives, i.e $P_i''\psi$, should vanish independently,
because the stems in bush are all different.
Thus, there cannot be more than one $\ov{A}$ per stem.
Next, we distribute one derivative per stem: each of them
cuts the $\ov{A}_{ab}$ from the stem leaving fixed indices at the end points,
leading to the expression,
\be
\left(\frac{\de\ \ }{\de \ov{A}_{ab}}\right)^2\Phi=
\eps_{cd}\ \ov{B}^{n_1}_{ca}\ \ov{B}^{n_2}_{da}\ \ 
\left(\ov{B}^{m_1}\psi\right)_{b} \left(\ov{B}^{m_2}\psi\right)_{b}\ .
\label{sol-app} 
\ee
This vanishes by antisymmetry of the epsilon tensor,
provided that $n_1=n_2$, i.e. that
the  $\ov{A}$ matrices appear at the same level on the two stems.
Furthermore, for $N>2$ one can repeat the argument, having $N-2$ spectator
stems over which the derivatives do not act; the height condition
should then applies for any pair of stems that have one $\ov{A}$.
In conclusion, all $\ov{A}$ should appear in the stems at the same height,
leading to the general solution (\ref{two-ll}).

\subsection{States obeying the $A^3=0$ projection}

We now discuss the solution of the $A^3=0$ condition.
Bushes can have one, two, three $\ov{A}$'s per stem and more: 
we consider each case in turn.
For three $\ov{A}$'s and more, $A^3=0$ can act on a single stem
and not vanish: this limits the number of $\ov{A}$'s per stem to two.

A) For bushes that have single-$\ov{A}$ stems only, we should examine
the action of $A^3$ an all triples of stems (1-1-1 action).
This vanishes by antisymmetry (cf. (\ref{sol-app})) 
if for any triple considered, two $\ov{A}$'s are at the same height.
It follows that on single-$\ov{A}$ bushes, the $\ov{A}$'s can stay
at two heights, i.e. form two bands.

B) For bushes with double-$\ov{A}$ stems only, $A^3$ can act
on pairs (2-1 action $(B_1)$) or on triples (1-1-1 action $(B_2)$)
of stems.

$B_1$) Consider the action 2-1 on the pair:
\ba
&\left(\frac{\de\ \ }{\de \ov{A}_{ab}}\right)^3 &
\eps_{ij} \left(C\ov{A}D\ov{A}E\right)_i \left(F\ov{A}G\ov{A}H\right)_j 
\nl
&&=\eps_{ij} \left[
\left(C_{ia} D_{ba} E_b\right)\left(F_{ja} G\ov{A}H_b\right) +
\left(C_{ia} D_{ba} E_b\right)\left(F\ov{A}G_{ja} H_b\right)
\right.
\nl
& & \ +\ 
\left.
\left(C_{ia} D\ov{A}E_b\right)\left(F_{ja} G_{ba} H_b\right) +
\left(C\ov{A}D_{ia} E_b\right)\left(F_{ja} G_{ba}H_b\right)
\right] \ .
\label{up-prot}
\ea
The first and third term in this equation vanish independently 
when $C=F$ due to the earlier identity $\eps_{ij}u_i u_j=0$;
the sum of the second and fourth term vanishes for $D=G$ due to
the possibility of factorizing an expression of the type, 
$\eps_{ij}\left( u_i v_j+ u_j v_i\right)=0$.
We thus found that the double-$\ov{A}$ stems should have $\ov{A}$
located on two heights (two bands).

$B_2$) There are $2^3=8$ possible actions 1-1-1 on triples of
stems involving two $\ov{A}$'s each. Having already enforced condition
($B_1$), their  $\ov{A}$'s are located on two bands.
The 8 terms generated by the action of $A^3$ 
are found vanish by the same two mechanism found in (\ref{up-prot}).
Therefore, there are no new conditions.

C) For bushes involving both double- and single-$\ov{A}$ stems,
we should again consider the actions 2-1 on pairs ($C_1$) and
1-1-1 on triples ($C_2$) of stems.

$C_1$) We consider the pair made by one double-$\ov{A}$ stem and 
one single-$\ov{A}$ stem; the double derivative acts necessarily on
the former stem, thus producing a unique term.
This vanishes as $\eps_{ij}u_i v_j=0$ if $u=v$, namely if 
the $\ov{A}$ on the single stem-$\ov{A}$ is located at the same height of 
the lower $\ov{A}$ in the double-$\ov{A}$ stem. 
It implies that the  $\ov{A}$ form again two
bands, but those on single-$\ov{A}$ stems should stay in the lower band.

$C_2$) The three derivatives act 1-1-1 on triples of stems with
number of   $\ov{A}$'s equal to (2,1,1) or (2,2,1),
yielding 2 and 4 terms respectively.
All these terms vanish independently, because
single-$\ov{A}$ stems already have their   $\ov{A}$ on the lowest band
by condition $C_1$.

In summary, the $A^3=0$ projection allows two $\ov{A}$ per stem at
most, that should form two bands. If both single- and double-$\ov{A}$
stems are present in the same bush, 
the  $\ov{A}$ on single stems should stay on the
lower band. All these features have been checked on the computer
for small-N examples.

\subsection{States with $A^4=0$ projection}

Again the action of the four derivatives on a single stem is
not vanishing and requires three $\ov{A}$ per  stem at most.
Hereafter we list the possible actions of the four derivatives.

A) If there are single-$\ov{A}$ stems only, the derivative
action is 1-1-1-1: for every four-plet of stems, two $\ov{A}$
should be at the same height; thus, three bands of $\ov{A}$
can be formed on bushes.

B) If there are double-$\ov{A}$ stems only, there can be:
($B_1$) 2-2 action on pairs of stems; ($B_2$) 2-1-1 action on 
triples of stems, ($B_3$) 1-1-1-1 action on four-plets of stems.

$B_1$) There is a single term that vanishes if the lower  $\ov{A}$
are at the same height.

$B_2$) There are 12 terms that vanish by the same two mechanisms
of $B_1$ in the previous $A^3=0$ case, provided the upper $\ov{A}$ 
form another band, i.e. stay at the  same height.

$B_3$) All terms vanish once the previous conditions are enforced.

In summary, double-$\ov{A}$ stems should have their $\ov{A}$'s on
two bands.

C) If there are triple-$\ov{A}$ stems only, there can be:
($C_1$) 3-1 and 2-2 actions on pairs; ($C_2$) 2-1-1 action on triples;
($C_3$) 1-1-1-1 on four-plets.

$C_1$) There are 6 terms for the 3-1 action and 9 for the 2-2 action:
these cancel individually or in pairs by the two mechanisms of $B_1$
 in the previous $A^3=0$ case, provided that all $\ov{A}$'s form
three bands.

$C_2$) The action 2-1-1 on 3-$\ov{A}$ stems generates 81 terms, 
that are satisfied once $C_1$ has been enforced.

$C_3$) The terms generated by the 1-1-1-1 action vanish because 
there are at least two derivatives of $\ov{A}$ at the same height.

In summary, triple-$\ov{A}$ stems should have their $\ov{A}$'s on
three bands.

D) Consider now the  case of stems having two or one $\ov{A}$ each,
as for states filling the second and third Landau level.
From the previous analysis we know that
the double-$\ov{A}$ form 2 bands (case (B)) and the triple-$\ov{A}$ stems
can have 3 bands (case (C)). We should only consider the new cases when 
the four derivatives act on stems of mixed type.
There can be:
($D_{11}$) 2-1-1 action on 2-2-1 stems; 
($D_{12}$) 2-1-1 action on 2-1-1 stems; 
($D_{21}$) 1-1-1-1 action on 2-2-2-1 stems; 
($D_{22}$) 1-1-1-1 action on 2-2-1-1 stems; 
($D_{23}$) 1-1-1-1 action on 2-1-1-1 stems; 

$D_{11}$) Given that one derivative acts on the single-$\ov{A}$
stem, the remaining three derivatives cancel as in case $(B_1)$ of $A^3=0$,
on stems already having two $\ov{A}$ bands. No new conditions.

$D_{12}$) The condition is that on any pair of single-$\ov{A}$ stems,
one of them has the $\ov{A}$ on the lowest band of the double-$\ov{A}$ stems.
This allows single-$\ov{A}$ to stay on any of the two bands, with some 
exceptions.

$D_{21}$) It is satisfied.
$D_{22}$) It yields the same condition as ($D_{12}$).
$D_{23}$) For every triple of single-$\ov{A}$ stems, two should be
on the same band. The solution is
that each of the 2 bands of double-$\ov{A}$ stems are allowed
(weaker than ($D_{12}$)).

In summary, mixed double- and single-$\ov{A}$ stems should have their
$\ov{A}$'s forming two bands with some exceptions.

E) The most relevant case for Jain's ground state at $\nu^*=4$ is
for mixed stems with one, two and three  $\ov{A}$'s.
Owing to the previous conditions, each individual type is already
organized in 3, 2 and 3 bands respectively.
The possible new actions of the four derivatives are the following ones:
($E_{11}$) 3-1 and 2-2 actions on pairs of type 3-2;
($E_{12}$) 3-1 action on pairs of type 3-1; 
($E_{21}$) 2-1-1 action on triples of type 3-3-2, 
($E_{22}$) on triples 3-2-2, ($E_{23}$) on triples 3-2-1  and 
($E_{24}$) on triples 3-1-1 ;
($E_4$) 1-1-1-1 actions on all stem types.

$E_{11}$) There is cancellation by the usual two mechanisms
($(B_1)$ of $A^3=0$)
provided that the $\ov{A}$'s of double-$\ov{A}$ stems stay in the
lowest of the three bands of triple-$\ov{A}$ stems.

$E_{12}$) As before, the $\ov{A}$ of single-$\ov{A}$ stems should all
align on the lowest of the three bands of the triple-$\ov{A}$ stems.

Once these two conditions are enforced, the other E-type actions
are checked.
In summary, mixed triple-, double- and single-$\ov{A}$ stems should have their
$\ov{A}$'s forming three bands, with the condition that stems with
less that three $\ov{A}$'s should align their $\ov{A}$'s 
on the lowest available
bands. This is the condition enforcing the uniqueness of the state
with maximal filling $\nu^*=4$ as explained in section 3.4.
The same mechanism works for the $A^m=0$ ground states with $\nu^*=m$
(\ref{m-jain}) that contain stems of any number of $\ov{A}$'s.


\end{document}